\documentclass[honours]{unswthesis}
\usepackage[top=2.5cm, bottom=2.5cm, left=2.5cm, right=2.5cm]{geometry}
\usepackage{bbm}
\usepackage{amsfonts}
\usepackage{amssymb}
\usepackage{amsthm}
\usepackage{latexsym,amsmath}
\usepackage{amsmath}
\usepackage{graphicx}
\usepackage{epstopdf}
\usepackage{fancyvrb}
\usepackage{float}
\usepackage{appendix}
\usepackage{algorithm}
\usepackage[noend]{algorithmic}
\algsetup{indent=2em}
\usepackage{mathrsfs}
\usepackage{multirow}
\usepackage{boites}
\usepackage{psfrag}
\usepackage{mathtools}
\usepackage{tikz}
\usetikzlibrary{trees}
\tikzstyle{every node}=[draw=black,thick,anchor=west]
\tikzstyle{selected}=[draw=red,fill=red!30]
\tikzstyle{optional}=[dashed,fill=gray!50]

\usepackage{fancyhdr}
\usepackage{multicol}
\usepackage{array}
\usepackage{color}
\DeclareMathOperator*{\argmin}{\arg\!\min}
\DeclareMathOperator*{\argmax}{\arg\!\max}

\DeclareMathOperator{\di}{d\!}

\newcommand{\Pm}{\mathbb{P}}
\newcommand{\simiid}{\stackrel{\mathrm{iid}}{\sim}}

\newcommand{\Var}{\mathrm{Var}}

\newcommand{\bX}{\mathbf{X}}
\newcommand{\bx}{\mathbf{x}}
\newcommand{\by}{\mathbf{y}}

\newcommand{\gvn}{|}
\newcommand{\Em}{\mathbb{E}}
\newcommand{\scX}{\mathscr{X}}
\newcommand{\scT}{\mathscr{T}}
\newcommand{\g}{\leftarrow}
\newcommand{\bLam}{\boldsymbol\Lambda}

\newcommand{\bb}[1]{\mathbb{#1}}
\renewcommand{\v}[1]{\boldsymbol{#1}}
\newcommand{\m}[1]{\mathrm{#1}}
\newcommand{\idef}{\stackrel{\mathrm{def}}{=}}

\DeclareGraphicsExtensions{.eps,.pdf,.png,.jpg}

\newtheorem{Theorem}{Theorem}[section]

\newtheorem{example}{Example}[chapter]
\title{The Dynamic Splitting Method with an application to portfolio
credit risk}

\authornameonly{Kevin Lam}

\author{\Authornameonly\\{\bigskip}Supervisor: Dr. Zdravko Botev}

\copyrightfalse
\figurespagefalse
\tablespagefalse

\begin{document}

\beforepreface

\declaration


\prefacesection{Acknowledgements}
I would like to sincerely thank my supervisor Dr. Zdravko Botev for his guidance, tutelage and intellectual discussions he has provided throughout my Honours year. I am grateful for him taking time to share his interesting research ideas that has perked my interests. Also, I must express my gratitude to my colleagues for accompanying me on the same journey in 2015. I would also like to thank the School of Mathematics and Statistics at UNSW for the enriching experience over the last 5 years. Finally, I would like to thank my family for their encouragement and support.

\prefacesection{Abstract}
We consider the problem of accurately measuring the credit risk of a portfolio consisting of loss exposures such as loans, bonds and other financial assets. We are particularly interested in the probability of large portfolio losses. We describe the popular models in the credit risk framework including \emph{factor models} and \emph{copula models}. To this end, we revisit the most efficient probability estimation algorithms within current copula credit risk literature, namely importance sampling. We illustrate the workings and developments of these algorithms for large portfolio loss probability estimation and quantile estimation. We then propose a modification to the dynamic splitting method which allows application to the credit risk models described. Our proposed algorithm for the unbiased estimation of rare-event probabilities, exploits the quasi-monotonic property of functions to embed a static simulation problem within a time-dependent Markov process. A study of our proposed algorithm is then conducted through numerical experiments with its performance benchmarked against current popular importance sampling algorithms.

\vfill
 



\afterpreface


%
%


\chapter{Introduction}
Simulation from intractable multidimensional distributions and estimates of corresponding real-valued quantities are hallmark problems in Monte Carlo methods; (see \cite{kroese2011handbook}). In this chapter we examine these two related problems more closely as they frequently arise in Monte Carlo applications.
The first problem is to simulate from the conditional density in $\mathbb{R}^d$
\begin{equation}
\label{cond}
f^*(\bx)= \frac{1}{\ell} f(\bx)\bb I\{S(\bx)\geq\gamma\},\quad \bx=(x_1,\ldots,x_d)^\top,
\end{equation}
where we assume that:    $S: \mathbb{R}^d\rightarrow\mathbb{R}$ is a real-valued function, which we refer to as an \emph{importance function}; $f$ is a density function on $\mathbb{R}^d$ such that $X_1,\ldots,X_d$ are independent; $\gamma$ is a real parameter, and 
\begin{equation}
\label{ell}
\ell=\Pm_f(S(\bX)\geq\gamma),\quad \bX\sim f
\end{equation} is a normalizing constant. $S$ can be interpreted as a measure of performance of $\bX$ with $\gamma$ being a loss threshold which, if exceeded, triggers an event whose associated probability we wish to estimate. To estimate the probability of this event, we must also solve a second problem, and that is to estimate the normalizing constant $\ell$ accurately and efficiently. $\ell$ can be interpreted the probability of default for a financial institution or a probability of ruin for an insurance company if $S(\bX)$ is the aggregate losses or claims incurred.

Note that, despite the specification of \eqref{cond}, the definition of the conditional density function $f^*$ comes as a special case of any complex high-dimensional density function $\tilde f(\bx)=\breve f(\bx)/\mathcal{Z}$ on $\mathbb{R}^d$  with a known or unknown normalizing constant $\mathcal{Z}$. In other words, without loss of generality,  the conditional density \eqref{cond} includes many models arising in Bayesian inference and statistics, econometrics and finance.

In the unlikely case that $\ell$ is not a \emph{rare-event probability} (that is, it is not too small; say, larger than $10^{-4}$), one can simulate  from \eqref{cond} exactly with the   acceptance-rejection algorithm. That is, we simulate $\bX\sim f$ until $S(\bX)\geq\gamma$ is satisfied. For the simulated $\bX$ for which this condition is satisfied, we accept this as a sample from the conditional distribution \eqref{cond}. Unfortunately, more often than not, $\ell$  is a rare-event probability  and in such cases the only practicable  approach  to  simulate from \eqref{cond} is approximate  Markov Chain Monte Carlo (MCMC) sampling as described in \cite{brooks2011handbook}.

In this thesis, we utilize Monte Carlo methods to estimate these probabilities of the form \ref{ell}. In particular, we focus on a classical Monte Carlo technique, Dynamic Splitting (DS), (see \cite{kahn1951estimation,kroese2011handbook}). In the original formulation of dynamic splitting, the state space of a Markov process is decomposed into nested subsets so that the rare event is represented as the intersection of sequentially decreasing event subsets.  Within each subset the sample paths of the Markov process are split into multiple copies; the rationale behind this is to promote and capture more occurrences of the rare event. As a result, the probability of the rare event is calculated as the product of conditional probabilities. Note that by splitting the Markov process into more copies, it allows more accurate estimation of each conditional probability.

The remaining chapters of the thesis are organized as follows. In Chapter~\ref{chapter: e-intro}, we cover essential background knowledge of Monte Carlo methods and describe algorithms for the efficient estimation of $\ell$, namely \emph{Importance Sampling} and \emph{Cross-Entropy}. In Chapter~\ref{chapter: c-intro}, we introduce and survey the framework of copula credit risk models in current literature with worked examples of the algorithms presented in Chapter \ref{chapter: e-intro}. Further, in Chapter 4, we illustrate the workings of our proposed algorithm through numerical experiments benchmark its performance against existing robust Monte Carlo estimators. Moreover, critical analysis of advantages and limitations of the proposed DS algorithm is
given. Finally, in Chapter 5, we provide concluding remarks and directions for future research. To preserve the flow of the thesis, we delegate verifications and proofs of key results to the appendix. 
\chapter{Background on Monte Carlo methods}
\label{chapter: e-intro}
The distribution of losses $F_L$ is the structure of interest in credit risk modelling. There is often no closed form for $F_L$ making direct calculations of $\ell(l)$ for a given $l$, or $l$ for a given $\ell$. To estimate these measures we require a method to draw independent and identically distributed (iid) samples of $L$ from $F_L$ and a method to estimate probabilities and quantiles given an iid sample $L_1, \dots , L_N$, particularly those in the upper tail of $F_L$ such as the Value-at-Risk (VaR) and Conditional Value-at-Risk (CVaR). The purpose of this chapter is to describe the estimators of probabilities and quantiles using Monte Carlo methods, namely \emph{Importance Sampling}. 

\section{Efficiency}
In this section, we describe the idea of efficiency as we require a criteria to benchmark performance of estimators illustrated in this paper. For a rare-event estimator, this is often measured by its \emph{relative error} (RE)\cite{kroese2011handbook}. This is the normalized standard deviation of the estimator. Suppose we have an unbiased estimator $\hat{\ell}$ of the rare-event probability defined as
\[
\begin{split}
\ell(\gamma) & = \bb P(S(\bX)\geq \gamma)\\
&=\bb E[\bb I\{S(\bX)\geq \gamma\}]
\end{split}
\]
where $f$ is a probability density function (pdf),$S$ is a real-valued function, $\bX$ is a random vector and $\gamma$ is a threshold parameter. The event of interest is $\{S(\bX)\geq \gamma\}$ occurring under $f$. RE is defined as 
\[
RE(\hat{\ell})=\frac{\sqrt{\Var(\hat{\ell})}}{\ell\sqrt{N} }
\]
where $N$ is the number of iid estimates of $\ell$.
\section{Crude Monte Carlo}
\emph{Crude Monte Carlo} (CMC) \cite{van2000asymptotic} approximates the cumulative distribution function (cdf) $F_L$ with the empirical distribution function (edf) $\hat{F}_L$. Let $L_1, \dots , L_N$, be an iid sample of size $N$, then $\hat{F}_L$ defined as follows
\[
\hat{\ell} (\gamma) = \frac{1}{N} \sum_{k=1}^N \bb I (L_k \geq \gamma).
\]

We define $\hat{v}_\alpha$ as the CMC estimate of the $\alpha$-VaR or equivalently the $\alpha$-quantile. By using the above definition, CMC sets $\hat{v}_\alpha$ as the solution to the problem
\[
\hat{v}_\alpha = \inf \{\gamma:\hat{\ell}(\gamma) \leq 1-\alpha\}.
\]

As a result, we can also obtain $\hat{v}_\alpha$ by sorting the $\{L_k\}$ in ascending order and taking the $\lceil \alpha N\rceil$-th largest value. Once $\hat{v}_\alpha$ is known, we can calculate the CMC estimator for the CVaR as follows
\[
\hat{c}_\alpha = \frac{1}{N(1-\alpha)} \sum_{k=1}^N L_k \bb I (L_k \geq \hat{v}_\alpha).
\]

It is well-known that the VaR of a portfolio is not additive, that is, it is not the sum of the VaR of sub-portfolios or individual losses as it is not the sum of independent random variables. Nevertheless,\cite{hong2014review} provides some insight into the asymptotic distribution of these estimators through the following central limit theorems.

\begin{Theorem}[Central Limit Theorems for CMC estimators]
If $L$ is positive and continuously differentiable density $f_L$ around $v_\alpha$ and $\bb E [L^2] < \infty$, then as $N\rightarrow \infty$ 
\[
\sqrt{N}(\hat{v}_\alpha - v_\alpha) \xrightarrow{d} \mathsf{N}\left(0,\frac{\alpha(1-\alpha)}{f_L(v_\alpha)^2}\right),
\]
\[
\sqrt{N}(\hat{c}_\alpha - c_\alpha) \xrightarrow{d} \mathsf{N}\left(0,\frac{\Var(L\bb I (L>v_\alpha))}{(1-\alpha)^2}\right).
\]
\end{Theorem}

\section{Importance Sampling}
Our aim is to efficiently estimate $\ell(\gamma)$ for a given $\gamma$, or $\gamma$ for a given $\ell$. CMC estimators heavily rely on the computational power of a large sample size to achieve accuracy. However, if $\ell$ is a rare-event probability say, smaller than $10^{-4}$, then generating large samples is costly to simulate large values of $\gamma$. Much of the work on simulation methods has been done on variance reduction algorithms to improve the efficiency of the CMC estimator. We now present an algorithm that is particularly well-suited to rare-event problems, \emph{Importance Sampling} (IS). IS has been shown to accurately and efficiently sample from the upper tail of a general loss distribution, and hence reduce the variance of $\hat{\ell}$ \cite{glasserman2005importance}.

Suppose that $L$ can be simulated under another density $g_L$, which we denote as the IS density. Define,
\[
W(\gamma) = \frac{f_L(\gamma)}{g_L(\gamma)}
\]
as the likelihood ratio obtained by the change in probability measures. Note that 
\begin{align*}
\bb P(L > \gamma) &= \bb E_f[\bb I \{L > \gamma\}]\\
&= \bb E_g [W(L) \bb I \{L > \gamma\}]
\end{align*}
where $\bb E_f$ and $\bb E_g$ denote expectations with respect to pdf $f_L$ and $g_L$ respectively. As we are interested in the upper tail of loss distribution, we can thus estimate tail probabilities as follows
\[
\hat{\ell} = \frac{1}{N} \sum_{k=1}^N W(L_k) \bb I \{L_k > \gamma\}.
\]
\cite{glynn1996importance} suggests the IS approach to quantile estimation by defining the IS cdf as
\[
\begin{split}
\hat{F}_L^{IS} (\gamma) &= 1 - \hat{\ell} \\
&=1-\frac{1}{N} \sum_{k=1}^N W(L_k) \bb I \{L_k > \gamma\}.
\end{split}
\]

With this definition, the IS estimators for the VaR and CVaR are given by
\[
\hat{v}_\alpha^{IS} = \inf \{\gamma\hat{F}_L^{IS} \geq \alpha\},
\]
\[
\hat{c}_\alpha^{IS} = \frac{1}{N(1-\alpha)} \sum_{k=1}^N W(L_k)L_k \bb I \{L_k \geq \hat{v}_\alpha^{IS}\}.
\]

Similar to the CMC estimators, \cite{hong2011monte} show that the IS estimators asymptotically follow Normal distributions and give the following central limit theorems.

\begin{Theorem}[Central Limit Theorems for IS estimators]
If $L$ is positive and continuously differentiable density $f_L$ around $v_\alpha$ and there exists $\epsilon>0$ and $p>2$ such that $W(\gamma)$ is bounded for all $\gamma \in (v_\alpha-\epsilon,v_\alpha+\epsilon)$ and $\bb E_g [\bb I (L\geq v_\alpha-\epsilon)(W(L))^p]<\infty$, then as $N\rightarrow \infty$ 
\[
\sqrt{N}(\hat{v}_\alpha^{IS} - v_\alpha) \xrightarrow{d} \mathsf{N}\left(0,\frac{\Var_{g}(W(L))\bb I (L\geq v_\alpha))}{f_L(v_\alpha)^2}\right),
\]
\[
\sqrt{N}(\hat{c}_\alpha^{IS} - c_\alpha) \xrightarrow{d} \mathsf{N}\left(0,\frac{\Var_{g}(W(L))L\bb I (L>v_\alpha))}{(1-\alpha)^2}\right).
\]
\end{Theorem}
where $\Var_g$ denotes the variance under the pdf $g_L$. Note that $g_L$ is not known so the above theorems only describe attractive properties of a good choice of $g_L$. A good choice for the IS pdf depends on the distribution of $\v X$, properties of the set $\{S(\bX)\geq\gamma\}$ and more importantly, the tail behaviour of $S(\bx)$. A light-tailed random variable $X$ is defined as one which has a finite moment generating function (mgf), that is $\bb E[e^{\theta X}]<\infty$ for $\theta>0$. \cite{brereton2013monte} suggests that a good IS pdf $g_L$ in a light-tailed setting is an exponentially twisted pdf derived from $f_L$, defined as 
\[
g_L(\gamma) = \frac{\exp (\theta \gamma)f_L(\gamma)}{\bb E[e^{\theta L}] }.
\] 
Here $\bb E[e^{\theta L}]$ acts as the normalizing constant to ensure $g_L$ is a density. The likelihood ratio of an exponentially twisted pdf is thus given by
\begin{equation}
\label{LLR}
W(\gamma) = E[e^{\theta L}]\exp (-\theta \gamma).
\end{equation}

Attractive properties of likelihood ratios of this form are described in \cite{bucklew2013introduction,dembo2009large}.

\subsection{Adaptive Importance Sampling}
Adaptive importance sampling methods aim to avoid theoretical complications and computational issues in rare-event probability estimation by deriving parameters for an optimal IS density by using sub-samples of sample data. As described previously, there is often no closed form for $f_L$ so we represent the portfolio loss in the form $L=S(\bX)$ and seek to parameterize a prespecified IS density. As this is discussed within a credit risk framework, our interest is to simulate from a density conditional on the event  $\{S(\bX)\geq \gamma\}$ where $\gamma$ is often chosen to be large loss threshold. If an initial sample $\bX_1,\ldots,\bX_M$ can be generated directly from the optimal IS density, that is, the zero-variance density $g^*(\bx)=f(\bx|S(\bx)\geq \gamma)$ then the parameters can be computed to approximate $g^*$. A popular and versatile adaptive importance sampling method is \emph{Cross-Entropy} (CE)\cite{rubinstein2011simulation,rubinstein2013cross}. The goal of CE is to specify a density $g$ `close' to $g^*$ so that both would behave similarly and give reasonably accurate IS estimators. We consider the family of distributions $\mathcal{G}=\{g(\v x; \v v)\}$ where $\v v$ denotes a vector of parameters. A convenient measure of the difference between two densities $g_1$ and $g_2$ is the \emph{Kullback-Leibler divergence} also known as the \emph{cross-entropy distance}:
\[
\mathcal{D}(g_1,g_2)=\int g_1(\bx)\log \frac{g_1(\bx)}{g_2(\bx)}\mathrm{d}\bx.
\]
Every density in $\mathcal{G}$ can be represented as $g(\cdot;\v v)$ for some $\v v$, so we obtain the optimal IS density by solving the following problem:
\[
\v v^*_{CE}=\argmin_{\v v} \mathcal{D}(g^*,g(\cdot;\v v)).
\]

This is equivalent \cite{rubinstein2011simulation} to solving 
\[
\v v^*_{CE}=\argmax_{\v v} \bb E[ f(\bX)\bb I\{S(\bX)\geq \gamma\}\log g(\bX;\v v)].
\]

This problem does not typically have an explicit solution. Instead we can estimate $\v v^*_{CE}$ by solving the following problem:
\[
\hat{\v v}^*_{CE}=\argmax_{\v v} \frac{1}{N} \sum_{i=1}^N \bb I\{S(\bX_i)\geq \gamma\}\log g(\bX_i;\v v)]
\]
where $\bX_1,\ldots.\bX_N$ are simulated from $f$. This is further simplified if we are able to draw approximately from $g^*$, reducing the problem to 
\[
\hat{\v v}^*_{CE}=\argmax_{\v v} \frac{1}{N} \sum_{i=1}^N \log g(\bX_i;\v v)]
\]
where $\bX_1,\ldots.\bX_N$ are drawn approximately from $g^*$.
Suppose we want to sample approximately from the zero-variance density $g^*(\bx)=f(\bx|S(\bx)>v_\alpha)$. Given a generated sample of portfolio losses $L_1,\ldots,L_N$ with corresponding vectors $\bX_{L_1},\ldots,\bX_{L_N}$ from $f$, we can order the portfolio losses in ascending order as $L_{(1)}\leq\cdots\leq L_{(N)}$ and choose $L_{(\lceil\alpha N\rceil)},\ldots,L_{(N)}$ with corresponding vectors $\bX_{L_{(\lceil\alpha N\rceil)}},\ldots,\bX_{L_{(N)}}$ as an approximate sample from $g^*$. With this approximate sample, we can use standard maximum likelihood estimation to parameterize a prespecified density $g$ which approximates $g^*$. The approximate sampling algorithm can be summarized as follows.
\begin{algorithm}[H]
\label{alg:zerovariancesampling}
\caption{: Sampling approximately from $g^*$}
\begin{algorithmic}
\REQUIRE{distribution of $\bX$ $f$; importance function $S$; sample size $N$; loss threshold $x$}
\STATE{$t\leftarrow0$}
\REPEAT 
  \STATE{$t\leftarrow t + 1$}
  \STATE{Simulate $\bX_t$  from $f$}  
  \STATE{$B_k\leftarrow \bb I\{X_k> x_k\}$ for $k=1,\ldots,d$}
  \STATE{Generate the portfolio loss $L_t \leftarrow S(\bX_t)$}
\UNTIL{$t = N$}
\STATE{Sort the losses $\v L\leftarrow (L_{[1]},\ldots,L_{[N]})^\top$}
\RETURN{$L_{(\lceil\alpha N\rceil)},\ldots,L_{(N)}$}
\AND{$\bX_{L_{(\lceil\alpha N\rceil)}},\ldots,\bX_{L_{(N)}}$} 
\end{algorithmic}
\end{algorithm}
\chapter{Introduction to Copula Credit Risk Models}
\label{chapter: c-intro}

\section{Summary of the Model}
The model can be summarized as follows.
Let
\[
L=c_1 B_1+\cdots+c_d B_d= \v c^\top  \v B
\]
be the total loss incurred by a portfolio of $d$ obligors, where
$c_k$ is the loss incurred from the $k$-th obligor and $B_i\sim \mathsf{B}(P_k)$ is a Bernoulli random variable indicating whether the $k$-th obligor has defaulted. The distribution of the column vector $\v B$ is implicitly defined under factor models and copula models which are described later in this chapter. For now, we note that dependence of default events is captured in the model through the default probabilities $\{P_k\}$ which contain a set of common factors, say $\Psi$, that affect all obligors. Conditional on $\Psi$, the problem is simplified to modelling the sum of independent Bernoulli variables $\{B_k\}$ scaled by the losses $\{c_k\}$. These models have popular applications in finance \cite{bluhm2010introduction,frey2003dependent,glasserman2003monte}, particularly in the valuation of credit risk such as the estimation of Value-at-Risk for a given confidence level $\alpha$. Popular values for $\alpha$ are $0.95$, $0.99$ and $0.995$ since the loss values of interest lie in the upper tail of the loss distribution. The purpose of this chapter is to describe the most popular credit risk models, \emph{factor models} and \emph{copula models}, with applications of the algorithms presented in Chapter~\ref{chapter: e-intro}. 

\section{Factor Models}
In factor models, $\v B$ are defined as
\[
B_k\idef \bb I\{X_k>x_k\},\quad k=1,\ldots,d
\]
where $\{x_k\}$ are given fixed thresholds and $\v X$ has a continuous joint density $f(\v x)$. 
For example, $\v X\sim\mathsf{N}(\v \mu, \m\Sigma)$ for some mean vector $\v \mu$ and covariance matrix $\m\Sigma$. Note that $\m \Sigma$ can be singular. 

\subsection{Gaussian factor model}
A popular Gaussian factor model for $\v X$ is 
\begin{equation}
\label{GFM}
X_k=a_{k1} Z_1+\cdots+a_{km} Z_m+b_k \epsilon_k,\quad k=1,\ldots,d
\end{equation}
where
$Z_1,\cdots,Z_m\simiid \mathsf{N}(0,1)$ are the so called systematic risk factors which affect all obligors;
$a_{k1},\cdots,a_{kd}$ are default factor loadings for the $k$-th obligor with 
\[
a_{k1}^2+\cdots+a_{km}^2\leq 1,
\]
\[
b_k=\sqrt{1-(a_{k1}^2+\cdots+a_{km}^2 )}.
\]
and $\epsilon_k\sim\mathsf{N}(0,1)$ is risk specific to the $k$-th obligor. We thus have the marginal distribution  $X_k\sim \mathsf{N}(0,1)$. We can write 
\[
\v X= \m A \v Z +\mathrm{diag}(\v b )\v \epsilon ,
\]
where $\m A$ is an $d\times m$ matrix.

Here, probabilities are conditionally independent on $\v Z$, that is $\Psi = \v Z$, the default probability of the $k$-th obligor $P_k(\v Z)$ is
\begin{equation}
\label{Gcon}
\begin{split}
P_k(\v Z) &= \bb P (X_k >x_k|\v Z=\v z)\\
&=\bb P \left(\epsilon_k >\frac{ x_k-(a_{k1} z_1+\cdots+a_{km} z_m)}{b_k}\right)\\
&=\Phi \left( \frac{(a_{k1} z_1+\cdots+a_{km} z_m)-x_k}{b_k}\right).
\end{split}
\end{equation}

\subsubsection{Crude Monte Carlo}
Suppose that the factor loadings matrix  $\m A$ and default thresholds $\v x$ are known. For a chosen sample size $N$ and confidence level $\alpha$, the CMC algorithm is as follows.
\begin{algorithm}[H]
\label{alg:GaussianCMC}
\caption{: Generating  $L$ under a Gaussian factor model using CMC}
\begin{algorithmic}
\REQUIRE{factor loadings  $\m A$; cost vector $\v c$; default thresholds $\v x$; sample size $N$; confidence level $\alpha$; loss threshold $\gamma$}
\STATE{$t\leftarrow0$}
\REPEAT 
  \STATE{$t\leftarrow t + 1$}
  \STATE{Simulate $\v Z \sim\mathsf{N}(\v 0,\m I_m)$}
  \STATE{Simulate $\v\epsilon\sim\mathsf{N}(\v 0,\m I_d)$}
  \STATE{$\v X\leftarrow \m A \v Z +\mathrm{diag}(\v b )\v \epsilon$}  
	\STATE{$B_k\leftarrow \bb I\{X_k> x_k\}$ for $k=1,\ldots,d$}
  \STATE{$L_t \leftarrow \v c^\top \v B$}
\UNTIL{$t = N$}
\STATE{Sort the losses $\v L\leftarrow (L_{[1]},\ldots,L_{[N]})^\top$}
\RETURN{$\hat{\ell} (\gamma) \leftarrow \frac{1}{N} \sum_{k=1}^N \bb I (L_k > \gamma)$} 
\OR{$\hat{v}_\alpha \leftarrow L_{\lceil \alpha \times N\rceil}$}
\end{algorithmic}
\end{algorithm}

\subsubsection{Importance Sampling}
Suppose that the conditional default probabilities $\v P(\v Z)$ are known, \cite{glasserman2005importance} describes an IS algorithm which changes the default indicators to $B_k \sim \mathsf{B}(P_k(\v Z))$ before applying exponential twisting to the conditional probabilities $P_k(\v Z)$ as follows 

\begin{equation}
\label{exptwist}
P_{k,\theta}(\v Z)=\frac{P_k(\v Z) e^{\theta c_k }}{1+P_k(\v Z) (e^{\theta c_k}-1)}.
\end{equation}

By applying this change in probability measure, conditional default probabilities are increased if $\theta>0$. Exponential twisting are well known for the significant variance reduction in range of contexts \cite{asmussen2007stochastic}. It is difficult to apply exponential twisting directly on $f_L$ so it is applied on the default indicators $B_k$ instead and this can be shown to give equivalent results \cite{glasserman2005importance}. Conditional on $\Phi=\v Z$, $L$ becomes a sum of independent scaled Bernoulli random variables $c_k B_k$. Now each $c_k B_k$ has mgf 
\[
\bb E[e^{\theta c_k B_k}]=(1-P_k(\v Z))+P_k(\v Z)e^{\theta c_k } <\infty
\]
so $\{c_k B_k\}$ are light-tailed with a good IS pdf being an exponential twisted density obtained by applying \eqref{exptwist}.  It can easily be verified that this leads to the likelihood ratio for $L$ given by \eqref{LLR}.  The algorithm, which \cite{glasserman2005importance} calls the one-step algorithm, can be summarized as follows. Given $\v Z$ and $\{x_k\}$, we calculate the conditional probabilities $P_k(\v Z)$ and apply exponential twisting to each probability. This generates an exponential twisted density for each random variable $c_k B_k$; the product of these densities forms the exponentially twisted density $g_L$. The likelihood ratio is then used to estimate upper tail probabilities from which quantiles in the tail can then be estimated. The one-step IS algorithm is as follows.

\begin{algorithm}[H]
\label{alg:glasserman1}
\caption{: Generating $L$ using one-step IS algorithm}
\begin{algorithmic}
\REQUIRE{factor loadings  $\m A$; cost vector $\v c$; default thresholds $\v x$; sample size $N$; confidence level $\alpha$; loss threshold $\gamma$}
\STATE{$t\leftarrow0$}
\REPEAT 
  \STATE{$t\leftarrow t + 1$}
  \STATE{Simulate $\v Z \sim\mathsf{N}(\v 0,\m I_m)$}
  
  \STATE{Calculate vector $\v P(\v Z)$ with $P_k (\v Z) \leftarrow \Phi\left(\frac{\sum_i \m A_{ki} Z_i-x_k}{b_k}\right)$}
  \IF{$\v c^\top \v P(\v Z) \geq \gamma$} 
  \STATE{$\theta \leftarrow 0$} 
  \ELSE 
  \STATE{ $\theta$ is assigned to be the solution to
	\[
	\gamma=
	\sum_{k=1}^m\frac{P_k(\v Z) c_k\exp(\theta c_k)}{1+P_k(\v Z) (\exp(\theta c_k) -1)}
	\]} 
  \ENDIF
  \FOR{$k=1,\ldots,d$}
	\STATE{$P_{k,\theta}(\v Z)\leftarrow \frac{P_k(\v Z) \exp(\theta c_k)}{1+P_k(\v Z)(\exp(\theta c_k)-1)}$}
	\STATE{Simulate $B_k\sim \mathsf{B}(P_{k,\theta}(\v Z))$, independently}
	\ENDFOR

  \STATE{Set $L_t \leftarrow \v c^\top \v B$  and
  \[ 
	W(L_t) \leftarrow  \exp\left(-\theta L_t+\sum_{k=1}^m \log\left(1+P_k(\v Z)(\exp(\theta c_k)-1)\right) \right)
	\]
  }

\UNTIL{$t=N$}
\RETURN{$\hat{\ell} (\gamma)\leftarrow \frac{1}{N} \sum_{t=1}^N \bb I \{L_t > \gamma\} W(L_t) $} 
\OR{$\hat{v}_\alpha^{IS} \leftarrow \inf \{\gamma:\hat{F}_L^{IS} (\gamma) \geq \alpha\}$}
\end{algorithmic}
\end{algorithm}

The efficiency of the algorithm depends on the level of dependence between obligors. \cite{glasserman2005importance} states that when dependence is weak, such as the case of the Gaussian factor model, increasing conditional default probabilities by exponential twisting already reduces variance in the estimators of tail probabilities and quantiles. The algorithm aims to simulate large values for $L$ centred around a carefully chosen loss threshold $\gamma$ that lies in the tail of $f_L$. This is done by solving for the unique value of $\theta$ as described in the algorithm and applying the exponential twisting with this value of $\theta$. An attractive property of exponential twisted pdf $g_L$, that can be easily verified, is that for each simulated $L$ we have
\[
\bb E_g[L|\v Z] \geq \gamma,
\] 
which comes a result of solving for $\theta$ such that
\[
\gamma=\sum_{k=1}^m\frac {P_k(\v Z) c_k\exp(\theta c_k)}{1+P_k(\v Z) (\exp(\theta c_k) -1)}.
\] Hence, the algorithm can  sample from the tail of $f_L$ if we chose our loss threshold $\gamma$ to be a quantile in the tail, such as $\hat{v}_{0.95}$, which can be initially estimated with CMC. With this sampling,  small portfolio losses are now rare events while large portfolio losses are now frequent.

The vector of likelihood ratios, $\v W$, can be used to estimate upper tail probabilities.  After sorting the simulated losses in ascending order, the IS estimator $\hat{v}_\alpha^{IS}$ can be computed as $\mathrm{VaR}=L_{(j)}$ where $j$ is the solution to
\[
\min \limits_j \left(\frac{1}{N}\sum_{k=j}^N W(L_{(k)}) \leq 1 - \alpha \right)
\]

\cite{glasserman2005importance} states a further extension to form a two-step algorithm. This is motivated by the variance decomposition of the estimator $\hat{\ell}$
\[
\Var(\hat{\ell})= \bb E\left(\Var(\hat{\ell|\v P})\right)+\Var\left(\bb E(\hat{\ell}|\v P)\right).
\]
The one-step IS algorithm minimizes the variability of $\hat{\ell}$, that is, it minimizes $\Var(\hat{\ell|\v P})$. The two-step IS algorithm aims to minimize $\Var\left(\bb E(\hat{\ell}|\v P)\right)$. This is equivalent to minimizing the variance of the CMC estimator $\hat{q}$ \cite{brereton2013monte} of
\[
q = \bb P(L>\gamma|\v P(\v Z)).
\]
The corresponding zero-variance density $g^*$ \cite{kroese2011handbook} is given by
\[
g^*_{\v Z}(\v z) \propto \bb P(L>\gamma|\v P(\v Z))f_{\v Z}(\v z).
\]
However, we must note that the normalizing constant $\ell$ is the same constant we wish to estimate so this is not a practical IS density. Nevertheless, this provides a direction in the searching for a good IS density. A common approach to  by applying the one-step algorithm after a change of measure for $\v Z$. In particular, we change the mean of $\v Z$. The underlying rationale of this approach is to generate more defaults by shifting the mean of the factors $\v Z$ by increasing each of its components.  This leads to high values for the default factor loadings $\{X_k\}$ which are more likely to exceed the thresholds $\{ x_k\}$. The challenge in the two-step algorithm is to describe a suitable IS distribution $g_Z$. \cite{glasserman2005importance} and \cite{glasserman1999asymptotically} propose using a Normal distribution $\mathsf{N}(\v \mu, \m I_m)$ with the same mode as optimal pdf $g_Z^{*}$. The mode $\v \mu^*$ is also the mean of the Normal distribution and provided that we have loss threshold $x$, is given as the solution to the following problem

\begin{equation}
\label{shiftfactorsoptim}
\v \mu^* = \argmax \limits_{\v z} \mathbb{P}\left(L>\gamma|\v Z = \v z \right)\exp {(-\frac{1}{2}\v z^\top \v z)}.
\end{equation}

Hence the two-step algorithm applies a change in distribution to the factors $\v Z$ and simulates $\v Z \sim \mathsf{N}(\v \mu^*,\m I_m)$. The difficulty now lies in solving \eqref{shiftfactorsoptim}. \cite{glasserman2005importance} states several approximations to simplify this problem. We focus on the \emph{constant approximation} and the \emph{tail bound approximation} as it provides the most convenient way to combine IS applied on the probabilities $P_k(\v Z)$ with IS applied on the factors $\v Z$. For details on other approximations used in solving \eqref{shiftfactorsoptim}, interested readers may refer to \cite{glasserman2005importance}.

The \emph{constant approximation} involves replacing $L$ with $\bb E[L|\v Z = \v z]$ and $\mathbb{P} (L>\gamma|\v Z = \v z)$ with $\bb I(\bb E[L|\v Z = \v z]>\gamma)$. This approximation replaces $\mathbb{P}\left(L>\gamma|\v Z = \v z \right)$ with a constant and so \eqref{shiftfactorsoptim} becomes
\begin{equation}
\label{constapprox}
\argmin \limits_{\v z} \{\v z^\top \v z:  \bb E[L|\v Z = \v z]>\gamma \}.
\end{equation}

The \emph{tail bound approximation} is an approach which aims to approximate $\mathbb{P}\left(L>\gamma|\v Z = \v z \right)$ by its upper bound. It then proceeds by maximising this upper bound which in turn, could maximise the probability $\mathbb{P}\left(L>\gamma|\v Z = \v z \right)$. Using this approximation, \eqref{shiftfactorsoptim} becomes 
\begin{equation}
\label{tailbound}
\argmax \limits_{\v z} \left\{\sum_{k=1}^m \log[1+P_k(\v Z)(\exp(\theta c_k)-1)]-\theta \gamma-\frac{1}{2}\v z^\top \v z\right\}.
\end{equation}
The two-step algorithm for dependent obligors is as follows.

\begin{algorithm}[H]
\label{alg:glasserman2}
\caption{: Generating $L$ using Glasserman and Li's two-step algorithm}
\begin{algorithmic}
\REQUIRE{factor loadings  $\m A$; cost vector $\v c$; default thresholds $\v x$; sample size $N$; confidence level $\alpha$; loss threshold $\gamma$; shifted mean vector $\v \mu^*$}
\STATE{$t\leftarrow0$}
\REPEAT 
  \STATE{$t\leftarrow t + 1$}
  \STATE{Simulate $\v Z \sim\mathsf{N}(\v \mu ,\m I_m)$}
  
  \STATE{Calculate vector $\v P(\v Z)$ with $P_k (\v Z) \leftarrow \Phi\left(\frac{\sum_j \m A_{k,j} Z_j-x_k}{b_k}\right)$}
  \IF{$\v c^\top \v P(\v Z) \geq \gamma$} 
  \STATE{$\theta \leftarrow 0$} 
  \ELSE 
  \STATE{ $\theta$ is assigned to be the solution to
	\[
	\gamma=
	\sum_{k=1}^m\frac{P_k (\v Z) c_k\exp(\theta c_k)}{1+P_k (\v Z) (\exp(\theta c_k) -1)}
	\]} 
  \ENDIF
  \FOR{$k=1,\ldots,d$}
	\STATE{$ P_{k,\theta} (\v Z)\leftarrow \frac{P_k (\v Z) \exp(\theta c_k)}{1+P_k (\v Z)(\exp(\theta c_k)-1)}$}
	\STATE{$B_k\leftarrow \mathsf{Ber}(P_{k,\theta})$, independently}
	\ENDFOR

  \STATE{Set $L_t \leftarrow \v c^\top \v B$  and
  \[
  W(L_t) \leftarrow  \exp\left(-\theta L+\sum_{k=1}^m \log\left(1+P_k (\v Z)(\exp(\theta c_k)-1)\right)\right)\exp \left(-\v\mu^{*\top} \v Z + \frac{\v \mu^{*\top} \v \mu^*}{2} \right)
  \]
  }
\UNTIL{$t=N$}
\RETURN{$\hat{\ell} (\gamma) \leftarrow \frac{1}{N} \sum_{t=1}^N \bb I (L_t > \gamma) W(L_t) $} 
\OR{$\hat{v}_\alpha^{IS} \leftarrow \inf \{\gamma:\hat{F}_L^{IS}(\gamma) \geq \alpha\}$}
\end{algorithmic}
\end{algorithm}

\subsection{$t$ Factor Model}
The $t$ factor model \cite{chan2010efficient} differs from the Gaussian factor model as the factors now have a multivariate $t$ distribution rather than a multivariate Normal distribution. Following the desired properties and notation from \eqref{GFM}, $\v X$ generated from a $t$ factor model usually has the representation
\[
X_k= \sqrt{\frac{r}{V}} \left( a_{k1} Z_1+\cdots+a_{km} Z_m+b_k \epsilon_k \right) ,\quad k=1,\ldots,d
\] where $Z_1,\cdots,Z_m\simiid \mathsf{N}(0,1)$ and $V\sim \mathsf{\chi}^{2} (r)$. Note $X_k$ is in the form 
\[
X_k=\frac{Z}{\sqrt{V/r}},
\]
where $Z\sim \mathsf{N}(0,1)$ and $V\sim \mathsf{\chi}^{2} (r)$. This implies that we have the marginal distribution $X_k\sim \mathsf{T} (r)$, which we denote as a $t$ distribution with $r$ degrees of freedom.

Here, probabilities are conditionally independent on $\v Z = \v z$ and $V=v$, that is $\Psi = \{\v Z,V\}$, the default probability of the $k$-th obligor $P_k(\v Z,V)$ is
\begin{equation}
\label{tcon}
\begin{split}
P_k(\v Z,V) &=\bb P (X_k >x_k|\v Z=\v z, V=v)\\
&= \bb P \left(\epsilon_k >\frac{ \sqrt{\frac{v}{r}} x_k-(a_{k1} z_1+\cdots+a_{km} z_m)}{b_k}\right)\\
&=\Phi \left( \frac{(a_{k1} z_1+\cdots+a_{km} z_m)-\sqrt{\frac{v}{r}}x_k}{b_k}\right).
\end{split}
\end{equation}

\subsubsection{Crude Monte Carlo}
Suppose that the factor loadings matrix  $\m A$ and default thresholds $\v x$ are known. For a chosen sample size $N$, confidence level $\alpha$ and degrees of freedom $r$, the CMC algorithm is as follows
\begin{algorithm}[H]
\label{alg:tCMC}
\caption{: Generating  $L$ under a $t$ factor model using CMC}
\begin{algorithmic}
\REQUIRE{factor loadings  $\m A$; cost vector $\v c$; default thresholds $\v x$; sample size $N$; confidence level $\alpha$; loss threshold $\gamma$; degrees of freedom $r$}
\STATE{$k\leftarrow0$}
\REPEAT 
  \STATE{$k\leftarrow k + 1$}
  \STATE{Simulate $V \sim\mathsf{\chi}^2(r)$}
  \STATE{Simulate $\v Z \sim\mathsf{N}(\v 0,\m I_m)$}
  \STATE{Simulate $\v\epsilon\sim\mathsf{N}(\v 0,\m I_d)$}
  \STATE{$\v X\leftarrow \sqrt{\frac{r}{V}}\left(\m A \v Z +\mathrm{diag}(\v b )\v \epsilon\right)$}  
	\STATE{$B_i\leftarrow \bb I\{X_i> x_i\}$ for $i=1,\ldots,d$}
  \STATE{$L_k \leftarrow \v c^\top \v B$}
\UNTIL{$k = N$}
\STATE{Sort the losses $\v L\leftarrow (L_{[1]},\ldots,L_{[N]})^\top$}
\RETURN{$\hat{\ell} (\gamma) = \frac{1}{N} \sum_{k=1}^N \bb I (L_k > \gamma)$} 
\OR{$\hat{v}_\alpha\leftarrow L_{\lceil \alpha \times N\rceil}$}
\end{algorithmic}
\end{algorithm}

\subsubsection{Importance Sampling}
From the well-founded IS framework under the Gaussian factor model, much of the literature consider exponential twisting to the $t$ factors by first conditioning on $V$. Under $t$ factor models we note that the factors $\v Z$ contribute little to the occurrence of defaults as opposed to the case of Gaussian factor models. Rather, it is the value of $V$ that plays a much bigger role as a common shock factor. \cite{kang2005fast} applies exponential twisting to $V$, with the twisting parameter $\theta_V$ found as the solution to a linearly constrained optimization problem. The approach makes use of the observation that, conditional on $V$, one can apply the same IS algorithm as in \cite{glasserman2005importance} with modified thresholds for each obligor. By asymptotically optimal results, it forms an approximate zero-variance IS pdf for $V$ and implicitly applies the \emph{constant approximation} through the constrained sets specified in the optimization problem when solving for the twisting parameter $\theta_V$. The algorithm is however, computationally expensive as it requires solving the optimization problem multiple times per sample. \cite{kang2005fast} suggests combining stratified sampling with IS to reduce this cost.

For a general single factor model where $V$ need not follow a Chi-squared or Gamma distribution, \cite{bassamboo2008portfolio} presents two IS algorithms which apply exponential twisting on $V$ and $W=\frac{1}{V}$ with the twisting parameter $\theta$ found by solving optimization problems on the uniform upper bound of the likelihood ratio estimator in a similar manner to the \emph{tail bound approximation}. Another recent advancement applies a shift in factors as described in the two-step IS algorithm and adjusting the degrees of freedom of $V$ to maintain independence \cite{scott2015extensions,scott2015general}.

It can be shown that for $\lambda^2=\frac{V}{r}$ where  $V\sim \mathsf{\chi}^{2} (r)$ then $\lambda^2 \sim \mathsf{G}(\frac{r}{2},\frac{r}{2})$. \cite{chan2010efficient} utilizes ordered values of $\{ \frac{X_k}{x_k} \}$ and applies the cross-entropy method to efficiently sample large loss probabilities from a general $t$ copula $m$ factor model. Conditional on $\v Z$ and $\v \epsilon$, we can arrange the order statistics of $\{ \frac{X_k}{x_k} \}$ with corresponding costs $\{c_i\}$. The event $\{L>x\}$ occurs when $\lambda < \frac{X_{(i)}}{x_{(i)}}$ where $i=\min\{j:\sum_{k=j+1}^d c_{(k)}\leq \gamma\}$ so we can write
\[
\begin{split}
\bb P(L>x|Z,\v \epsilon)&=\bb P\left(\lambda < \frac{X_{(i)}}{x_{(i)}}\Bigg|Z,\v \epsilon\right)\\
&=F_\mathsf{G} \left(\frac{X_{(i)}^2}{x_{(i)}^2}\Bigg|Z,\v \epsilon\right).
\end{split}
\]
With this formulation, the cross-entropy method is applied to choose an optimal IS density $g^*(\v Z, \v \epsilon;\v v^*)$ which belongs to the parametric density family $\mathcal{F}$ defined as
\[
\begin{split}
\mathcal{F} =\left \{f(\v Z, \v \epsilon;\v v)=\prod_{j=1}^mf(Z_j; \mu_{Z},\sigma^2_{Z})\prod_{k=1}^d f(\epsilon_k; \mu_{\v \epsilon},\sigma^2_{\v \epsilon}) \right \}
\end{split}
\] 
where $\v v^*=(\mu^*_{Z},\m \sigma^{2*}_{Z},\mu^*_{\v \epsilon}, \sigma^{2*}_{\v \epsilon})$ and $\v v=(\mu_{Z},\m \sigma^{2}_{Z},\mu_{\v \epsilon}, \sigma^{2}_{\v \epsilon})$. This formulation gives the following IS estimator
\[
\hat{\ell}=\frac{1}{N} \sum_{k=1}^N F_\mathsf{G} \left(\frac{X_{(i)}^2}{x_{(i)}^2}\Bigg|\v Z^*_k,\v \epsilon^*_k\right) \frac{f(\v Z^*_k,\v \epsilon^*_k;\v v)}{g^*(\v Z^*_k,\v \epsilon^*_k;\v v^*)}.
\]

\subsection{Numerical Example: Gaussian and $t$ Factor Models}
We illustrate CMC and IS with an example from \cite{kroese2011handbook} and \cite{glasserman2005importance} where we apply the two-step IS algorithm for the Gaussian and CE for the $t$ factor model. Here we consider a portfolio of size $d =1000$ under a $m=21$ factor model with costs, marginal probabilities and thresholds as follows.
\[
\begin{split}
c_k &=\left( \lceil \frac{5k}{d} \rceil \right)^2,\\
P_k &= 0.01 \times \left( 1 + \sin \left( \frac{16\pi k}{d}\right)\right),\\
x_{\mathsf{G},k} &= \Phi^{-1}(1-P_k),\\
x_{\mathsf{T}(r),k} &= F^{-1}_{\mathsf{T}(r)}(1-P_k), \quad k=1,\ldots,d
\end{split}
\]
where $x_{\mathsf{G},k}$ and $x_{\mathsf{T}(r),k}$ are the default thresholds for the $k$-th obligor under the Gaussian and $t$ factor model with $r=3$ degrees of freedom respectively.
The factor loadings matrix $\m A$ has the block structure
\[ 
\m A = 
 \begin{pmatrix}
 \v r
 \begin{bmatrix}
 \v f &  &  \\
 & \ddots &  \\
 &  & \v f
 \end{bmatrix}
 \begin{array}{c}
 \m G \\
 \vdots \\
 \m G
 \end{array}
 \end{pmatrix}, 
 \quad \textrm{with $\m G= 
 \begin{pmatrix}
 \v g &  &  \\
 & \ddots &  \\
 &  & \v g
 \end{pmatrix}$},
\] 
where $\v r$ is a column vector of $1000$ entries, all equal to $0.8$; $\v f$ is a column vector of $100$ entries, all equal to $0.4$; $\m G$ is a $100\times10$ matrix with $\v g$ a column vector of $10$ entries, all equal to $0.4$. The conditional probabilities used in the two-step IS algorithm are calculated as in \eqref{Gcon}.

\begin{table}[h!]
\centering
\caption{Estimation of risk measures for a Gaussian factor model}
\begin{tabular}{|c c c c c c c c|}

 \hline
 $\alpha$ & $N$ & $\hat{\ell}^{IS}$ & $RE(\%)$ & $\hat{v}_\alpha$ & $\hat{v}_\alpha^{IS}$ & $\hat{c}_\alpha$ & $\hat{c}_\alpha^{IS}$\\ [0.5ex] 
 \hline
 $0.95$ & $10^4$ & $0.0482$ & $1.08$ & $530$ & $548$ & $1607$ & $1650$\\ 
 $0.99$ & $10^5$ & $0.01$ & $0.59$ & $2310$ & $2361$ & $3720$ & $3862$\\
 $0.995$ & $10^5$ & $0.005$ & $0.63$ & $3376$ & $3039$ & $4863$ & $5585$\\ [1ex] 
 \hline
\end{tabular}

\label{table:1}
\end{table}

\begin{table}[h!]
\centering
\caption{Estimation of risk measures for a $t$ factor model}
\begin{tabular}{|c c c c c c c c|}

 \hline
 $\alpha$ & $N$ & $\hat{\ell}^{CE}$ & $RE(\%)$ & $\hat{v}_\alpha$ & $\hat{v}_\alpha^{CE}$ & $\hat{c}_\alpha$ & $\hat{c}_\alpha^{CE}$\\ [0.5ex] 
 \hline
 $0.95$ & $10^4$ & $0.0480$ & $1.26$ & $388$ & $352$ & $2144$ & $1934$\\ 
 $0.99$ & $10^5$ & $0.0106$ & $0.48$ & $2934$ & $3072$ & $4733$ & $5171$\\
 $0.995$ & $10^5$ & $0.0060$ & $0.77$ & $4272$ & $4684$ & $5931$ & $6539$\\ [1ex] 
 \hline
\end{tabular}

\label{table:2}
\end{table}
We first use CMC to generate a sample of size $N$ for our initial estimates $\hat{v}_\alpha$ and $\hat{c}_\alpha$. We proceed to apply the two-step IS algorithm where we set $\hat{v}_\alpha$ as the loss threshold $\gamma$. In effect, we aim to simulate around the $\alpha$-VaR for the significance level specified in the above table. To apply the two-step IS algorithm, we have used the tail bound approximation to find the shifted mean vector for the risk factors $\v \mu^*$. Note that the sample size has been increased for higher levels of $\alpha$ to allow for generation of more loss values in the upper tail of the loss distribution.
After obtaining our parameter estimates for the two-step IS algorithm and CE, we have run $10$ iterations of both methods; with each iteration generating an elite sample of size $10^4$ to calculate $\hat{\ell}^{IS}$ and $\hat{\ell}^{CE}$. The mean and relative error of the $10$ values of $\hat{\ell}^{IS}$ and $\hat{\ell}^{CE}$ are calculated with the results shown in Table \ref{table:1}. From Table \ref{table:1}, we can see that the mean of the values $\{\hat{\ell}^{IS}\}$ and $\{\hat{\ell}^{CE}\}$ are close to the desired values of $1-\alpha$ with RE of $0.59\%$ to $1.08\%$ and $0.48\%$ to $1.26\%$ respectively.

\section{Copula Models}
A copula is defined as a multivariate distribution in the form
\[
C\left(u_1, \dots, u_d \right) = \mathbb{P} \left(U_1 \leq u_1, \dots, U_d \leq u_d \right),
\]
where $U_1,\cdots ,U_d$ are marginal uniformly distributed variables. Now $U_1,\cdots ,U_d$ can be written with respect to random variables $X_1,\cdots,X_n$  with marginal distributions $F_1,\cdots,F_d$. Thus we have
\[
\left(U_1, \dots, U_d \right) = \left(F_1(X_1),\dots ,F_d(X_d) \right).
\]
Hence the dependency of $\{X_k\}$ can be described individually through their marginal distributions by setting
\[
\left(X_1, \dots, X_d \right) = \left(F_{1}^{-1}(U_1),\dots ,F_{n}^{-1}(U_d) \right).
\]
We focus our discussion on a popular class of copulas known as Archimedean Copulas. Archimedean copulas have the following form
\[
C\left(u_1, \dots, u_d \right) = \psi ^{-1} \left( \psi(u_1) + \cdots + \psi(u_d) \right),
\]
where the function $\psi : [0,1] \rightarrow [0,\infty]$ is strictly decreasing with $\psi (0)=\infty$, $\psi (1)=0$ and its inverse $\psi ^{-1}$ monotonic. This class of copulas includes the Gumbel copula, where $\psi_{\eta}(u)=\left(-\log u \right)^{\eta}$, and the Clayton copula, where $\psi_{\eta} (u) = \frac{1}{\eta} (u^{-\eta} - 1)$. Note that the Gumbel copula has dependence in the upper tail while the Clayton copula has dependence in the lower tail. 

In the case of the Archimedean copulas, we can simulate the requisite
vector $\v U=(U_1,\ldots,U_d)$ as follows. 
First, simulate $\Lambda\geq 0$ from the distribution $F_\Lambda(\lambda)$, where
the pdf $f_\Lambda$ has the Laplace transform
\[
\int_0^\infty  f_\Lambda(\lambda)\exp(-u\lambda) \m d \lambda=\psi^{-1}(u),\quad u\geq 0.
\]
Note that $\psi^{-1}(u)$, with $\psi^{-1}(0)=1$ and  $\psi^{-1}(\infty)=0$, is then a decreasing completely monotone function, as required for the copula definition.
Then, given $\Lambda$, simulate  $E_1,\ldots,E_n\simiid \mathsf{Exp}(1)$ and output
\[
(U_1,\ldots,U_d)=\left(\psi^{-1}\left(\frac{E_1}{\Lambda}\right),\ldots, \psi^{-1}\left(\frac{E_d}{\Lambda}\right)\right),
\]
\[
(X_1,\ldots,X_d)=\left(F_1^{-1}\left(\psi^{-1}\left(\frac{E_1}{\Lambda}\right)\right),\ldots, F_d^{-1}\left(\psi^{-1}\left(\frac{E_d}{\Lambda}\right)\right)\right),
\]

Here, probabilities are conditionally independent on $\Lambda = \lambda$, that is $\Psi = \Lambda$, the default probability of the $k$-th obligor $P_k(\Lambda)$ is
\[
\begin{split}
P_k(\Lambda) &= \bb P (X_k>x_k|\Lambda=\lambda)\\
&=\bb P (E_k<\lambda \psi(F_k(x_k))) \quad \textrm{ since $\psi$ is invertible and decreasing}\\
&=1-\exp (\lambda \psi(F_k(x_k))).
\end{split}
\]

where $\{x_k\}$ are fixed thresholds similar to those present in factor models. Note that the nature of thresholds in copula models are often default times \cite{li1999default} and although this is also possible to model under factor models, the thresholds under factor models are more easily interpreted as economic and market risk thresholds.

\subsection{Crude Monte Carlo}
Suppose that the distribution $F_\Lambda$ is known. For a chosen sample size $N$ and confidence level $\alpha$, the CMC algorithm is as follows

\begin{algorithm}[H]
\label{alg:ACMC}
\caption{: Generating  $L$ under a Archimedean copula model using CMC}
\begin{algorithmic}
\REQUIRE{cost vector $\v c$; default thresholds $\v x$, sample size $N$; confidence level $\alpha$; loss threshold $x$; distribution $F_\Lambda$}
\STATE{$t\leftarrow0$}
\REPEAT 
  \STATE{$t\leftarrow t + 1$}
  \STATE{Simulate $\Lambda \sim F_\Lambda$}
  \STATE{Simulate $E_i \sim\mathsf{Exp}(1)$ for $i=1,\ldots,d$}
  \STATE{$\v U\leftarrow \frac{\v E}{\Lambda}$}
  \STATE{$X_k\leftarrow F_k^{-1}(U_k)$ for $k=1,\ldots,d$}
  \STATE{$B_k\leftarrow \bb I\{X_k> x_k\}$ for $k=1,\ldots,d$}
  \STATE{$L_t \leftarrow \v c^\top \v B$}
\UNTIL{$t = N$}
\STATE{Sort the losses $\v L\leftarrow (L_{[1]},\ldots,L_{[N]})^\top$}
\RETURN{$\hat{\ell} (x) = \frac{1}{N} \sum_{k=1}^N \bb I (L_k > x)$} 
\OR{$\hat{v}_\alpha\leftarrow L_{\lceil \alpha \times N\rceil}$}
\end{algorithmic}
\end{algorithm}

\subsection{Importance Sampling}
Under an Archimedean copula model, $\{X_k\}$ are independent conditional on $\Lambda=\lambda$ with probabilities
\[
P_k(X_k>x_k) = 1-\exp (\lambda \psi(F_k(x_k))).
\]
Thus, we can apply the one-step IS algorithm similar to the case of the Gaussian factor model by now generating the default indicators as $B_k \sim \mathsf{B}(P_k(\Lambda))$. We then proceed by applying exponential twisting to the probabilities $P_k(\Lambda)$ with twisting parameter $\theta$, computing $L$ and the likelihood ratio
\[
W(L)=\exp\left(-\theta L+\sum_{k=1}^m \log[1+P_k(\v Z)(\exp(\theta c_k)-1)] \right)
\]
for a sample size $N$. Our IS estimator for $\ell$ remains unchanged
\[
\hat{\ell} (\gamma) = \frac{1}{N} \sum_{t=1}^N \bb I (L_t > \gamma) W(L_t).
\]

\subsection{Numerical Example: a Clayton Copula Model}
We now illustrate CMC and IS for a Clayton copula model with marginal $\mathsf{Exp}(1)$ factors. The copula parameter $\eta$, default indicators and costs are given as follows
\[
\begin{split}
\eta &=5.5,\\
B_k &= \bb I (X_k>3),\\
c_k &=1 \quad k=1,\ldots,d.
\end{split}
\]

\begin{table}[h!]
\centering
\caption{Estimation of risk measures for a Clayton copula model}
\begin{tabular}{|c c c c c c c c|}

 \hline
 $\alpha$ & $N$ & $\hat{\ell}^{IS}$ & $RE(\%)$ & $\hat{v}_\alpha$ & $\hat{v}_\alpha^{IS}$ & $\hat{c}_\alpha$ & $\hat{c}_\alpha^{IS}$\\ [0.5ex] 
 \hline
 $0.95$ & $10^4$ & $0.0476$ & $5.30$ & $55$ & $55$ & $94$ & $92$\\ 
 $0.99$ & $10^5$ & $0.0093$ & $2.92$ & $121$ & $119$ & $162$ & $158$\\
 $0.995$ & $10^5$ & $0.0041$ & $5.16$ & $154$ & $146$ & $197$ & $187$\\ [1ex] 
 \hline
\end{tabular}

\label{table:3}
\end{table}
We first use CMC to first generate a sample of size $N$ and used to give the estimates $\hat{v}_\alpha$ and $\hat{c}_\alpha$. We proceed to apply the one-step IS algorithm where we use $\hat{v}_\alpha$ as the loss threshold $\gamma$. In effect, we aim to simulate around the $\alpha$-VaR for the significance level specified in the above table. Note that the sample size has once again, been increased for higher levels of $\alpha$ to allow for generation of more loss values in the upper tail of the loss distribution. $10$ iterations of the one-step algorithm are then run; with each iteration generating an elite sample of size $10^4$ to calculate $\hat{\ell}^{IS}$. The mean and relative error of the $10$ values of $\hat{\ell}^{IS}$ are calculated with the results shown in Table \ref{table:2}. From Table \ref{table:2}, we can see that although the mean of the values $\{\hat{\ell}^{IS}\}$ are close to the desired values of $1-\alpha$, the relative error of the one-step IS algorithm is $2.92\%$ to $5.30\%$ as opposed to $0.59\%$ to $1.08\%$ from the two-step IS algorithm. From this, we can see that the shift in factors for the Gaussian factor does indeed provide greater variance reduction for $\{\hat{\ell}^{IS}\}$. We now wish to compare this performance with our proposed algorithm, which we refer to as the \emph{Dynamic Splitting Method}.
\chapter{Splitting Simulation}
We now assume that $S$ is a quasi-monotone function, that is, $x_i\leq y_i,\; i=1,\ldots,d$ implies $S(\bx)\leq S(\by)$. 
In other words, a quasi-monotone function is one in which an increase in one of its components cannot reduce its value. It turns out that a large number of applied models either possess this property, or can be transformed into models possessing it. In the absence of any special properties of $f$ or $S$ in \eqref{cond}, there is little hope that one can do better than MCMC.  

The purpose of this chapter is to show how we can simulate efficiently from \eqref{cond} whenever the importance function $S$ is a \emph{quasi-monotone}  function. $S$ is a quasi-monotone function if  $x_i\leq y_i,\; i=1,\ldots,d$, implies $S(\bx)\leq S(\by)$. In other words, a quasi-monotone function is one in which an increase in one of its components cannot reduce its value. It turns out that a large number of applied models either possess this property, or can be transformed into  models possessing it. 

The proposed algorithm is an ingenious application of  the DS algorithm for simulation of Markov processes conditional on a rare event \cite{kahn1951estimation}. In its original form, DS cannot be applied to simulate from \eqref{cond}, because there is no underlying Markov process that we can split.  Our algorithm can be viewed as a way of transforming the problem of simulation from \eqref{cond} to one which involves the simulation of a Markov process. The underlying idea is to embed the static density \eqref{cond} within a continuous time Markov process in such a way that, at a particular instant of time, the Markov process has the exact same density as \eqref{cond}. 
Given this embedding, we can then apply the original splitting method of \cite{kahn1951estimation}. 

We emphasize that, unlike  \emph{generalized  splitting}
\cite{botev2012efficient,botev2008efficient} and 
\emph{subset simulation} \cite{au2001estimation}, the classical splitting algorithm does not employ MCMC sampling. 

\section{Classical Splitting Method}
\label{classicsplit}
In classical splitting, we consider a Markov process $\{\bX_t,t\geq0\}$ with the importance function $S$ over a state space $\scX$. It is assumed that $S(\bX_0)=0$ and that for any threshold $\gamma>0$, there are unique entry times to the sets $\{S(\bX_t)\geq \gamma\}$ and $\{S(\bX_t)\leq 0\}$ respectively given by
\[
\tau_\gamma=\min\{t:S(\bX(t))\geq\gamma\},
\]
\[
\tau_0=\min\{t:S(\bX(t))\leq 0\}.
\]

Note that $\tau_0$ only exists in the absence of the quasi-monotonic property in $S$, which is the case in classical splitting. Here the probability of interest is $\ell=\bb P(\tau_\gamma <\tau_0)$; the probability that the process reaches the threshold $\gamma$ before reaching $0$. Hence, $\ell$ depends on the distribution of $\bX_0$. Suppose there exists thresholds $\gamma_1$ and $\gamma_2$ such that $\gamma_2>\gamma_1$. The cornerstone of the splitting method is the observation that the process must first reach $\gamma_1$ to reach $\gamma_2$, and thus, giving us a sequence of nested event subsets
\[
E_{\gamma_2} =\{ \tau_{\gamma_2} <\tau_0 \} \subset E_{\gamma_1} =\{ \tau_{\gamma_1} <\tau_0 \}.
\] 
Hence $\ell = \bb P(E_{\gamma_2}\gvn E_{\gamma_1})\bb P(E_{\gamma_1})$, a product of conditional probabilities. This can similarly be extended to more threshold levels, say $0=\gamma_0<\gamma_1\cdots<\gamma_L=\gamma$ giving us 
\[
E_{\gamma_0} \supseteq E_{\gamma_1}\supseteq \cdots \supseteq E_{\gamma_L}.
\] 
Let $c_i=\bb P(E_{\gamma_i}\gvn E_{\gamma_{i-1}})$ then we have $\ell=\prod_{i=1}^Lc_i$. Each $c_i$ is then estimated as follows. 

Define $\scX_i=\{\bX_{t_i}:S(\bX_{t_i})\geq \gamma_i\}$, that is, the set of states in the Markov process that reach threshold $\gamma_i$ at time $t_i$ for $t_1, \cdots ,t_L$. At each $t_{i} \in \{t_1, \cdots ,t_L \}$, we run $s_i$ copies of the Markov process $\{\bX_{t_{i-1}}\}$ for each $\bX_{t_{i-1}} \in \scX_{i-1}$; giving $s_i|\scX_{i-1}|$ copies of $\{\bX_{t_{i}}\}$ and the corresponding evolved process $\{S(\bX_{t_{i}})\}$. Each copy of $\{\bX_{t_{i-1}}\}$ is run until $\{S(\bX_{t_i})\}$ either reaches the set $\{S(\bX_{t_i})\geq \gamma_i\}$ or $\{S(\bX_{t_i})\leq 0\}$ with ending state $\bX_{t_{i}}$. Each state $\bX_{t_i}$ that reaches the set  $\{S(\bX_{t_i})\geq \gamma_i\}$ before the set $\{S(\bX_{t_i})\leq 0\}$, referred to as an entrance state, is then stored as the set $\scX_i$. It is from these states that the next iteration of the splitting method will begin from; giving rise to $s_{i+1}|\scX_i|$ sample paths. An unbiased estimate of $c_i$ is given by $\hat{c}_i=\frac{|\scX_i|}{s_i|\scX_{i-1}|}$. It is clear that $|\scX_i|$ is dependent on the entrance states $\scX_{i-1}$. Despite this dependence, \cite{asmussen2007stochastic,glasserman1998look,kroese2011handbook} notes that the following estimate remains unbiased
\[
\begin{split}
\hat{\ell}&=\prod_{i=1}^L\frac{|\scX_i|}{s_i|\scX_{i-1}|}\\
&=\frac{|\scX_{L}|}{|\scX_{0}|}\prod_{i=1}^L\frac{1}{s_i}.
\end{split}
\]
Note we begin splitting at $t_1$ rather than $t_0$. In the above framework, the integer-valued splitting factors $\{s_i\}$ are distinct and thus may vary for each $t_{i} \in \{t_1, \cdots ,t_L \}$. This need not be the case as a predetermined splitting factor $s_i=s$ for all $i$, may be used; this version of the algorithm is referred to as \emph{Fixed Factor Splitting}. We will use this in combination with \emph{Fixed Effort Splitting} described later for our results in Chapter \ref{chapter: numerical}. The splitting process is repeated for all $t_{i} \in \{t_1, \cdots ,t_L \}$ with $s_i|\scX_{i-1}|$ being the \emph{simulation effort} at $t_i$. Potential problems when using the splitting method are large growths in the simulation effort and inefficiency. These problems come from inappropriate choices in the number of levels $L$, intermediate threshold levels $\{\gamma_1,\ldots,\gamma_{L-1}\}$ and splitting factors $\{s_1,\ldots,s_L\}$. \cite{kroese2011handbook} notes that ideally, levels should be chosen in such a way that the conditional probabilities $\{c_i\}$ can be easily estimated with CMC. Under the assumption of independence between the computational cost and time from running the Markov process, the total simulation effort is a random variable with expected value
\[
\begin{split}
\sum_{i=1}^L s_i\bb E [|\scX_{i-1}|]&=\sum_{i=1}^L s_i N_0 \prod_{j=1}^{i-1}c_js_j\\
&=N_0\sum_{i=1}^L \frac{1}{c_i}  \prod_{j=1}^ic_js_j
\end{split}
\]

If $c_js_j>1$ for all $j$, the simulation effort would become large as it increases with the number of levels $L$. This phenomenon is referred to as an explosion in \cite{glasserman1998large, kroese2011handbook}. If $c_js_j<1$ for all $j$, most sample paths will not reach the threshold levels $\gamma_j$ and as a consequence, the algorithm will be inefficient. Thus, an ideal choice is $s_j=\frac{1}{c_j}$ for all $j$. An alternative approach to avoiding explosions is \emph{Fixed Effort Splitting}, where the simulation effort is fixed for all $t_{i} \in \{t_1, \cdots ,t_L \}$ to say N, giving us a corresponding estimator
\[
\hat{\ell}_{\mathrm{FE}}=\prod_{i=1}^L \frac{|\scX_i|}{N}.
\]

Our results in Section \ref{chapter: numerical} will be based on this approach. Now that a basic description of the classical splitting method is given, we present the workings of our proposed algorithm including the embedding of the static density $\eqref{cond}$ in the Markov process $\bX_t$.

\section{The Dynamic Splitting Method for Static Problems}
We first require a way to apply the classical splitting method of \cite{kahn1951estimation} to the problem of sampling from \eqref{cond}.  To achieve this, we induce an artificial Markov process whose paths we can then repeatedly split to encourage entry into a desired rare-event set, say $\{\bX: S(\bX)\geq\gamma\}$ . 

A suitable  Markov process for our algorithm is the multivariate L\'evy subordinator \cite{barndorff2001multivariate}. A $d$-dimensional  L\'evy subordinator $\{\bLam(t),t\in \mathbb{R}_+\}$ with $\bLam(0)=0$ is an almost surely increasing stochastic process on a probability space $(\Omega,\Pm,\mathcal{F})$ with a continuous index set on $\mathbb{R}_+$ and with a continuous state space 
$\mathbb{R}^d$  defined by the following  properties: 
 (1) the increments of $\{\bLam(t)\}$ are stationary and non-negative, that is,  $(\bLam(t+s)-\bLam(t))\geq \mathbf{0}$  has the same distribution as  $\bLam(s)\geq\mathbf{0}$ for all $t,s\geq 0$; we denote the density of this stationary distribution as $\nu_s(\v\lambda)$;
 (2) the increments of $\{\bLam(t)\}$ are independent, that is, $\bLam(t_i)-\bLam(t_{i-1}),\; i=1,2,\ldots$ are independent for any
 $0\leq t_0<t_1<t_2<\cdots$; and
 (3) for any $\epsilon>0$, we have $\Pm(\|\bLam(t+s)-\bLam(t)\|\geq\epsilon)=0$ as $s\downarrow 0$. The distributional properties of the L\'evy subordinator are characterized by the \emph{characteristic exponent} which is expressed as the logarithm of the characteristic function of the random column  vector $\bLam(1)$:
\renewcommand{\i}{\mathrm{i}}
\begin{equation*}
\log\Em[\exp(\i \v s^\top \bLam(1))]=\i \v s^\top\v\mu+\int_{\mathbb{R}^d}\left(\exp(\i \v s^\top  \bx)-1-\i \v s^\top \bx\,\bb I_{\{\|\bx\|\leq 1\}}\right) \nu(\di \bx)\;,\quad \v s\in\mathbb{R}^d,
 \end{equation*}
for some $\v\mu\in \mathbb{R}^d$, and measure $\nu$
such that $\nu(\{\mathbf{0}\})=0$ and  $\int_{\mathbb{R}^d}\min\{1,\|\bx\|^2\}\nu(\di \bx)<\infty$. 

One of the simplest multivariate subordinators  we can have  is the \emph{gamma process with independent components} \cite[Section 3.2]{barndorff2001multivariate} in which the components $\Lambda_1(t),\ldots,\Lambda_d(t)$  of vector $\bLam(t)=(\Lambda_1(t),\ldots,\Lambda_d(t))^\top$ 
are independent  one-dimensional subordinators with characteristic exponent
$\log((1-i s)^{-1})$. In other words, each $\Lambda(t)$ has gamma distribution with shape parameter $t$ and scale parameter $1$, which denote by $\mathsf{G}(t,1)$. Note that other multivariate subordinators are possible as long as they possess the quasi-monotonicity property, but the Gamma process with independent components will suffice for our illustrations.

\begin{example}[Simulating Gamma process]
Consider simulating a  Gamma process at distinct times. 

\begin{algorithm}[H]
\caption{: Simulating Gamma Process at distinct times}
\begin{algorithmic}
\REQUIRE { Intermediate times $0=t_0<t_1<t_2<\cdots<t_L=1$}
\STATE{$\bLam(t_0)\leftarrow 0$}
\FOR{$i=1,\ldots,L$}
\STATE {Simulate $\bLam^*(t_i-t_{i-1})$  independently from subordinator distribution.}

\STATE{ $\bLam(t_{i})\leftarrow \bLam(t_{i-1})+\bLam^*(t_i-t_{i-1})$}
\ENDFOR
\RETURN{$\v\Lambda(t_{1}),\ldots,\v\Lambda(t_{L})$}
\end{algorithmic}
\end{algorithm}

\begin{figure}[h]
\caption{Simulation of 1D Gamma processes}
\begin{center}
\includegraphics[scale=.6]{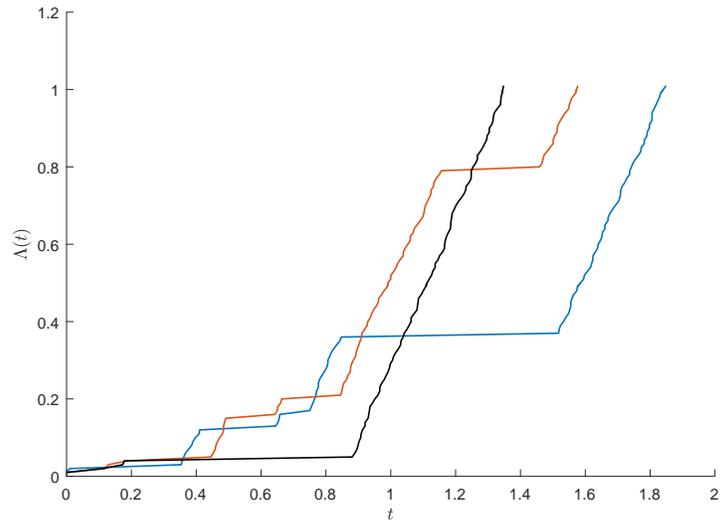}
\end{center}
\end{figure}

\begin{figure}[h]
\caption{Simulation of a 3D Gamma process}
\begin{center}
\includegraphics[scale=.8]{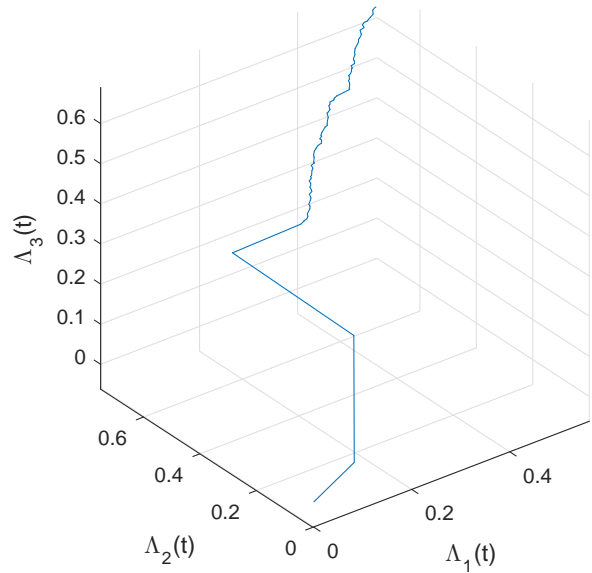}
\end{center}
\end{figure}

\end{example}
Now that we have defined one of the simplest continuous-time processes (the gamma subordinator above), 
we can proceed to embed the distribution of $\bX=(X_1,\ldots,X_d)^\top$ in \eqref{cond} within a 
continuous time process as follows. Let $F_k(x)$ be the cdf of $X_k$ for $k=1,\ldots,d$. Define the vector $\bX(t)=(X_1(t),\ldots,X_d(t))^\top$ through the random variables
\begin{equation}
\label{embedtransform-dec}
X_k(t)=F^{-1}_k(\exp(-\Lambda_k(t)))
\end{equation}
With this formulation, the non-decreasing property of the gamma subordinator implies that we will have an almost surely partial (and in fact total) ordering for the vectors $\bX(t)\succ\bX(t+s)$, meaning that $X_i(t)\geq X_i(t+s)$  for all $i$ and $s,t>0$. In addition, it can be easily verified that at time $t=1$ each $X_k(1)$ has the desired distribution $\Pm(X_k(1)\leq x)=F_k(x)$. This observation is what connects our induced Markov process to the static distribution of $\bX$. In fact, we can view the realization of the original vector $\bX$ as a snapshot of the state of a multivariate continuous-time process $\{\bX(t),t\geq 0\}$ at the instant $t=1$. Notice that $\Lambda_k(0)=0$ so we have $X_k(0)=\infty$. Consequently, we begin simulations in the set $\{S(\bX\geq\gamma)\}$ and evolve the gamma subordinator towards the set $\{S(\bX<\gamma)\}$ as a result of the ordering for the vectors $\bX(t)\succ\bX(t+s)$ for all $i$ and $s,t>0$.  As this property suggests, we will refer to \eqref{embedtransform-dec} as a \emph{monotonically decreasing embedding transformation}. Our numerical experiment results will be based on this transformation.

The above formulation now has a number of implications. First, the quasi-monotonicity of the importance function $S(\bX)$ implies that the exit time exists and is unique
\begin{equation}
\label{exit time}
\tau_\gamma=\sup\{t:S(\bX(t))\geq\gamma\}
\end{equation}
and in fact
$\Pm(S(\bX(t))\geq \gamma)=\Pm(\tau_\gamma>t).$
Second, if
$
0=t_0<t_1<\cdots<t_L=1
$
is any sequence of increasing times, the state space can be decomposed into the decreasing sequence of events:
 \[
\mathbb{R}^d\equiv \{S(\bX(t_0))\geq \gamma\}\supseteq \{S(\bX(t_1))\geq \gamma\}\cdots\supseteq \{S(\bX(t_{L}))\geq \gamma\}
\] 
and therefore we have the following decomposition of \eqref{ell}
\begin{align*}
\ell=\Pm(S(\bX(1))\geq \gamma)&=\prod_{i=1}^L\Pm(S(\bX(t_{i}))\geq \gamma \gvn S(\bX(t_{i-1}))\geq \gamma)\\
&=\prod_{i=1}^L\Pm(\tau_{\gamma}>t_i\gvn \tau_{\gamma}>t_{i-1})\,.
\end{align*}

At each intermediate point in time $t_{i} \in \{t_1, \cdots ,t_{L-1} \}$ for $i=1,\dots,L-1$, we consider $s$ splits in $s^{i-1}$ sample paths of the gamma subordinator for each component of $\bLam (t_i)$. From this, it is clear the estimation of each $\Pm(\tau_{\gamma}>t_i\gvn \tau_{\gamma}>t_{i-1})$ can be done empirically as
\[
\Pm(\tau_{\gamma}>t_i\gvn \tau_{\gamma}>t_{i-1})= \frac{|\scX_i|}{s|\scX_{i-1}|}.
\]
Note that since we begin the splitting of the sample paths at $t=t_1$ so we have 
\[
\Pm(\tau_{\gamma}>t_1\gvn \tau_{\gamma}>t_{0})= \frac{|\scX_1|}{|\scX_{0}|}.
\]
$\ell$ will thus be calculated as
\[
\begin{split}
\ell&=\frac{|\scX_1|}{|\scX_{0}|}\prod_{i=1}^{L-1}\frac{|\scX_{i+1}|}{s|\scX_{i}|}\\
&=\frac{|\scX_{L}|}{s^{L-1}|\scX_{0}|}\\
&=\frac{|\scX_{L}|}{s^{L-1}} \quad \textrm{if $|\scX_{0}|=1$}.
\end{split}
\]

For completeness, we will also define the \emph{monotonically increasing embedding transformation}. We consider the vector $\bX(t)=(X_1(t),\ldots,X_d(t))^\top$ with random variables is defined as 
\begin{equation}
\label{embedtransform-inc}
X_k(t)=F^{-1}_k(1-\exp(-\Lambda_k(t))).
\end{equation}

This transformation utilizes the observation that for any Uniformly distributed random variable $U$, $1-U$ is also uniform. This comes from the observation that $\Lambda_k(1) \sim \mathsf{Exp}(1)$ so its corresponding cdf value $1-\exp(-\Lambda_k(1))$ is Uniformly distributed and thus, $\exp(-\Lambda_k(1))$ is also Uniformly distributed. It can be easily verified that \eqref{embedtransform-inc} will also provide us with the desired distribution $\bb P(X_k(1)\leq x)=F_k(x)$. However, the consequences of this transformation differs from those of \eqref{embedtransform-dec}.

The non-decreasing property of the Gamma subordinator will now lead to increases in $X_k(t)$ as $t$ increases. As a result, we will now have the partial or total ordering for the vectors  $\bX(t)\prec\bX(t+s)$, that is, $X_i(t)\leq X_i(t+s)$ for all $i$ and $s,t>0$. As $\Lambda_k(0)=0$ we will have $X_k(0)=\min \{-\infty,0\}$ depending on the marginal distribution $F_k$. We will assume $X_k(0)=0$ for our purposes as we aim our discussion to the models described in Chapter~\ref{chapter: c-intro}. We note that all random variables that require this transformation in these models are indeed non-negative. Under this transformation we have $S(\bX(0))=0$ and begin the Markov process in the set $\{S(\bX(t))< \gamma\}$. We wish to estimate $\ell$ by modelling the entry time to the set $\{S(\bX(t))\geq \gamma\}$
\[
\tau_\gamma=\min\{t:S(\bX(t))\geq\gamma\},
\]

Hence, under the transformation \eqref{embedtransform-inc} we arrive at the classical splitting method as described in section \ref{classicsplit} with the modification that $\tau_0$ does not exist as a consequence of the quasi-monotonicity of the importance function and that we now seek
\begin{align*}
\ell = \bb P(S(\bX(1))\geq \gamma) &=\prod_{i=1}^L\Pm(S(\bX(t_{i}))\geq \gamma \gvn S(\bX(t_{i-1}))\geq \gamma)\\
&=\prod_{i=1}^L\Pm(\tau_{\gamma}<t_i\gvn \tau_{\gamma}<t_{i-1}).
\end{align*}

We can summarize the fixed factor dynamic splitting algorithm as follows.

\begin{figure}[h]
\caption{Splitting of paths}
\begin{center}
\psfrag{S}{$S$}
\psfrag{g}{$\gamma$}
\psfrag{t1}{$t_1$}
\psfrag{t2}{$t_2$}
\psfrag{1}{$1$}
\psfrag{0}{$0$}
\includegraphics[scale=.4]{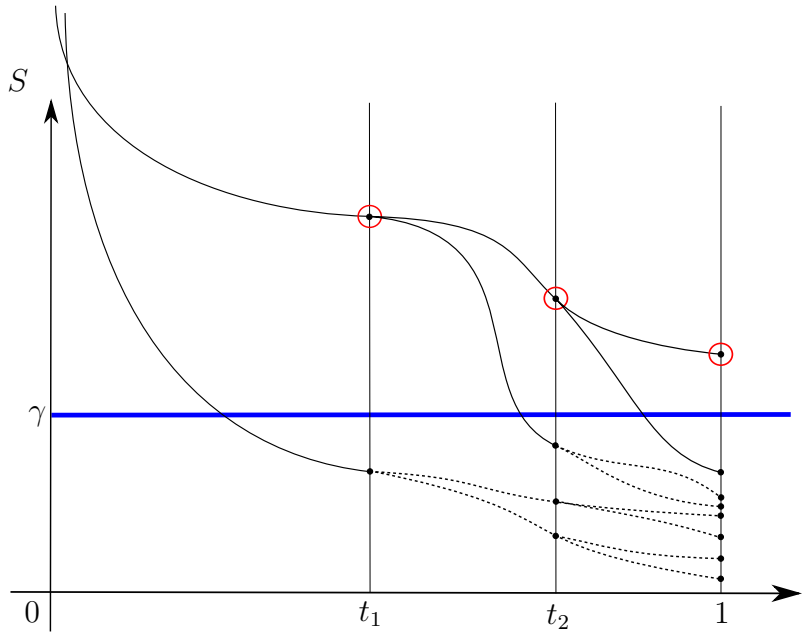}
\end{center}

\end{figure}

\begin{algorithm}[H]
\caption{: Fixed Factor Dynamic Splitting, returning $W$, an unbiased estimate of $\ell$}
\label{alg: dynamic splitting fixed factor}
\begin{algorithmic}
\REQUIRE {Splitting factor $s$ and intermediate times
 $0<t_1<t_2\cdots<t_L=1$}
 \STATE{Generate $\bLam(t_1)=(\Lambda_1(t_1),\ldots,\Lambda_{d}(t_1))^\top$ from the 
subordinator distribution.}
 \IF{$ S(\bLam(t_1)) >  \gamma$}
   \STATE{$\scX_1 \leftarrow \{\bLam(t_1)\}$}
 \ELSE
   \RETURN $W\g 0$
 \ENDIF 
 \FOR{$i=2,\ldots,L$}
\IF{$\scX_{i-1}=\emptyset$}
\RETURN $W\g 0$
\ELSE
   \STATE{$\scX_i \g \emptyset$}
   \FORALL {$\bLam(t_{i-1})\in\scX_{i-1}$}
     \FOR {$j=1,\ldots,s$}
       \STATE {For  $k=1,\ldots,d$ sample independently 
			\[
			\Lambda_k^*(t_i-t_{i-1})\sim\mathsf{G}(t_i-t_{i-1},1).
			\]
}
			\STATE{
			$\bLam^*(t_i-t_{i-1})\g(\Lambda_1^*(t_i-t_{i-1}),\ldots,\Lambda_d^*(t_i-t_{i-1}))^\top
			$
			}
			\STATE{ $\bLam(t_{i})\g \bLam(t_{i-1})+\bLam^*(t_i-t_{i-1})$}
       \IF{$ S(\bLam(t_i)) > \gamma$}
         \STATE{add $\bLam(t_i)$ to $\scX_i$}
       \ENDIF 
     \ENDFOR
   \ENDFOR
	\ENDIF
 \ENDFOR
 \RETURN {$W \g |\scX_{L}| / s^{L-1}$ as an unbiased  estimate.}
\end{algorithmic}
\end{algorithm}

In the above formulation, the splitting factor $s$ is chosen arbitrarily but under two idealizing assumptions, a near optimal value can be chosen for $s$ (see Appendix~\ref{optimsplit}).

As noted previously, \ref{alg: dynamic splitting fixed factor} has the risk of explosions so we implement a \emph{fixed effort} variant of the above algorithm to avoid this. At each $t_i$, we fix the simulation effort to $s$ splits for a randomly chosen $\bLam(t_{i-1}) \in \scX_{i-1}$. The fixed effort dynamic splitting algorithm can be summarized as follows.

\begin{algorithm}[H]
\caption{: Fixed Effort Dynamic Splitting, returning $W$, an unbiased estimate of $\ell$}
\label{alg: dynamic splitting fixed effort}
\begin{algorithmic}
\REQUIRE {Total sample at each level $s$; intermediate times
 $0<t_1<t_2\cdots<t_L=1$}
\STATE{$\scX_1\leftarrow\emptyset$}
\FOR{$k=1,\ldots,s$}
 \STATE{Generate $\bLam(t_1)=(\Lambda_1(t_1),\ldots,\Lambda_{d}(t_1))^\top$ from the 
subordinator distribution.}
 \IF{$ S(\bLam(t_1)) >  \gamma$}
   \STATE{Add $\bLam(t_1)$ to $\scX_1$}
 \ENDIF 
\ENDFOR
 \FOR{$i=2,\ldots,L$}
\IF{$|\scX_{i-1}|=0$}
\RETURN $W\g 0$
\ELSE
   \STATE{$\scX_i \g \emptyset$}
     \FOR {$j=1,\ldots,s$}
		\STATE{Let $\bLam(t_{i-1})$ be a randomly chosen member of $\scX_{i-1}$.}
       \STATE {For  $k=1,\ldots,d$ sample independently 
			\[
			\Lambda_k^*(t_i-t_{i-1})\sim\mathsf{G}(t_i-t_{i-1},1).
			\]
}
			\STATE{
			$\bLam^*(t_i-t_{i-1})\g(\Lambda_1^*(t_i-t_{i-1}),\ldots,\Lambda_d^*(t_i-t_{i-1}))^\top
			$
			}
			\STATE{ $\bLam(t_{i})\g \bLam(t_{i-1})+\bLam^*(t_i-t_{i-1})$}
       \IF{$ S(\bLam(t_i)) > \gamma$}
         \STATE{add $\bLam(t_i)$ to $\scX_i$}
       \ENDIF 
   \ENDFOR
	\ENDIF
 \ENDFOR
 \RETURN {$W \leftarrow \prod_{i=1}^L |\scX_i|/s^L$ as an estimate of $\ell$.}
\end{algorithmic}
\end{algorithm}

\section{Numerical Experiments}
\label{chapter: numerical}
We now illustrate the performance of dynamic splitting (DS) in the credit risk models described in Chapter~\ref{chapter: c-intro} through numerical experiments. For these results, we first verify that the importance function $S(\bX)=L(\bX)$ is indeed quasi-monotonic and specify suitable transformation \eqref{embedtransform-dec} and \eqref{embedtransform-inc} to embed the static random variables $\{X_k\}$ into time-dependent random variables $\{X_k(t)\}$. We first note that all of these models have the form
\[
\begin{split}
S(\bX)&=L(\bX)\\
&=\v c^\top  \v B(\bX)
\end{split}
\]

where the $k$-th component of $\v B$ is generated as $\bb I(X_k>x_k)$. An increase in one of the components of $\bX$ will lead to the value of $L(\bX)$ either staying the same or increased due to the activation of the corresponding indicator variable. It is clear that the function value of $L(\bX)$ will not be reduced so the importance function $S(\bX)=L(\bX)$ does indeed possess the quasi-monotonicity property required for our algorithm. We now present the results of the application of our algorithm to the \emph{Gaussian factor model}, \emph{$t$ factor model} and \emph{Clayton copula model}. In the following results, we have applied two-step IS for the Gaussian factor model, the CE-based estimator from \cite{chan2010efficient} for the $t$ factor model and one-step IS for the Clayton copula module. We note that the threshold $\gamma$ is chosen with an initial run of IS before our splitting algorithm is applied.

In the following results, $\gamma$ is the $\alpha$-quantile of the loss distribution estimated with IS from Chapter~\ref{chapter: c-intro} while $\hat{\ell}^{IS}$ and $\hat{\ell}^{DS}$ are the probabilities of the a loss exceeding $\gamma$. Hence, ideally we expect $\alpha+\hat{\ell}^{IS}$ and $\alpha+\hat{\ell}^{DS}$ to both equal to $1$. Disregarding the initial simulation effort for $\gamma$, all algorithms implemented below have been constrained to a simulation effort of $10^5$ and the same number of runs $R=10$. For IS and CE estimators, this is simply $NR$ where $N=10^4$ is the sample size in each run of the IS algorithm. For the DS estimator, it is $sTR$ where $s=1000$ is the splitting factor and $T=10$ is the number of intermediate time levels. Note, we have made appropriate adjustments to $s$ and $T$ to match the simulation effort of both algorithms in each comparison below.

\subsection{Factor Models}
We illustrate the performance of the splitting algorithm with factor models. We assume that $\m A$ has the matrix structure given in our worked example
\[ 
\m A = 
 \begin{pmatrix}
 \v r
 \begin{bmatrix}
 \v f &  &  \\
 & \ddots &  \\
 &  & \v f
 \end{bmatrix}
 \begin{array}{c}
 \m G \\
 \vdots \\
 \m G
 \end{array}
 \end{pmatrix}, 
 \quad \textrm{with $\m G= 
 \begin{pmatrix}
 \v g &  &  \\
 & \ddots &  \\
 &  & \v g
 \end{pmatrix}$},
\] 
where $\v r$ is a column vector of $1000$ entries, all equal to $0.8$; $\v f$ is a column vector of $100$ entries, all equal to $0.4$; $\m G$ is a $100\times10$ matrix with $\v g$ a column vector of $10$ entries, all equal to $0.4$. $\v b$ is calculated as 
\[
b_k=\sqrt{1-(a_{k1}^2+\cdots+a_{km}^2 )} 
\] 

\subsubsection{Gaussian Factor Model}
We have
\[
X_k=a_{k1} Z_1+\cdots+a_{km} Z_m+b_k\epsilon_k
\]
where $Z_1,\cdots,Z_m\simiid \mathsf{N}(0,1)$ and $\epsilon_k\sim \mathsf{N}(0,1)$. By using \eqref{embedtransform-dec} we have
\[
X_k(t)=a_{k1} \Phi^{-1}(\Lambda_1(t))+\cdots+a_{km} \Phi^{-1}(\Lambda_m(t))+b_k\Phi^{-1}(\Lambda_{m+k}(t)).
\]
Here we require gamma processes for $m$ standard Normal random variables which are the systematic risk factors $\v Z$ and $d$ standard Normal random variables for the obligor-specific risk factors $\v \epsilon$. 

\begin{table}[h!]
\centering
\caption{Estimation of $\ell$ for a Gaussian factor model}
\begin{tabular}{|c c c c c c|}

 \hline
 $\alpha$ & $\gamma$ & $\hat{\ell}^{IS}$ & $\hat{\ell}^{DS}$& $RE^{IS}(\%)$ & $RE^{DS}(\%)$\\ [0.5ex] 
 \hline
 $0.95$ & $548$ & $0.0493$ & $0.0497$ & $0.73$ & $4.35$\\ 
 $0.99$ & $2361$ & $0.0098$ & $0.0103$ & $0.60$ & $6.82$\\
 $0.995$ & $3039$ & $0.0062$ & $0.0056$ & $0.58$ & $7.28$\\ [1ex] 
 \hline
\end{tabular}

\label{table:4}
\end{table}

We can see that both $\alpha+\hat{\ell}^{IS}$ and $\alpha+\hat{\ell}^{DS}$ do indeed equal to $1$ approximately. However, the relative error of the DS estimator has a much larger relative error $RE^{DS}$ than the relative error for the two-step IS estimator $RE^{IS}$. Thus, in this study the DS estimator does not perform as efficiently as the two-step IS estimator.

\subsubsection{$t$ Factor Model}
We have
\[
\begin{split}
X_k&=\sqrt{\frac{r}{V}}\left(a_{k1} Z_1+\cdots+a_{km} Z_m+b_k\epsilon_k\right)\\
&=\sqrt{\frac{1}{G}}\left(a_{k1} Z_1+\cdots+a_{km} Z_m+b_k\epsilon_k\right)
\end{split}
\]
where $Z_1,\cdots,Z_m\simiid \mathsf{N}(0,1)$, $\epsilon_k\sim \mathsf{N}(0,1)$ and $G\sim \mathsf{G}(\frac{v}{2},\frac{v}{2})$. By using \eqref{embedtransform-dec} and \eqref{embedtransform-inc} we have
\[
X_k(t)=\frac{\left(a_{k1} \Phi^{-1}(\exp(-\Lambda_1(t)))+\cdots+a_{km} \Phi^{-1}(\exp(-\Lambda_m(t)))+b_k\Phi^{-1}(\exp(-\Lambda_{m+k}(t)))\right)}{\sqrt{F_\mathsf{G}^{-1}(1-\exp(-\Lambda_{d+m}(t)))}}.
\]
Here we require gamma processes for $m$ standard Normal random variables which are the systematic risk factors $\v Z$, $d$ standard Normal random variables for the obligor-specific risk factors $\v \epsilon$ and another for the common random variable $G$.

\begin{table}[h!]
\centering
\caption{Estimation of $\ell$ for a $t$ factor model}
\begin{tabular}{|c c c c c c|}

 \hline
 $\alpha$ & $\gamma$ & $\hat{\ell}^{CE}$ & $\hat{\ell}^{DS}$ & $RE^{CE}(\%)$ & $RE^{DS}(\%)$\\ [0.5ex] 
 \hline
 $0.95$ & $352$ & $0.0500$ & $0.0478$ & $0.36$ & $3.43$\\ 
 $0.99$ & $3072$ & $0.0100$ & $0.0096$ & $1.00$ & $6.55$\\
 $0.995$ & $4684$ & $0.0050$ & $0.0047$ & $0.72$ & $5.76$ \\ [1ex] 
 \hline
\end{tabular}

\label{table:5}
\end{table}

The performance of the CE estimator for the $t$ factor model has similar performance efficiency as that of the two-step IS estimator for the Gaussian factor model. This supports our theoretical motivations of these estimators since both estimators aim to significantly reduce the variance of $\hat{\ell}$ (see Chapter~\ref{chapter: e-intro}) where CE algorithm aims to sample approximately from the zero-variance density by learning near optimal parameters for the likelihood function while two-step IS reduces the variance via exponential twisting and change in parameter measures. Hence, we expect the relative error of both estimators to be quite small as supported by the above empirical results. We observe that $\alpha+\hat{\ell}^{IS}$ and $\alpha+\hat{\ell}^{DS}$ equal to $1$ approximately. However, the relative error of the DS estimator remains larger than the relative error for the CE estimator $RE^{CE}$. Thus, in this study the DS estimator does not perform as efficiently as the one-step IS estimator.

\subsection{Clayton Copula Model}
In this example, we consider the variables $\{X_k\}$ with marginal exponential distributions $\mathsf{Exp}(1)$. Note that $\{X_k\}$ are static variables computed as 
\[
\begin{split}
X_k &= F_k^{-1}(U_k)\\
&=-\log\left(1-\psi^{-1}\left(\frac{E_k}{G}\right)\right)
\end{split}
\]
where $F_k$ is the marginal distribution of $X_k$, $U_k$ is a uniform variable generated from the Clayton copula, $G\sim \mathsf{G}(\frac{1}{\eta},1)$ and $\psi^{-1}(t)=(1+\eta t)^{\frac{1}{\eta}}$. Let $F_\eta$ be the cdf of the $\mathsf{G}(\frac{1}{\eta},1)$ then an appropriate embedding transformation is
\[
X_k(t)=-\log\left(1-\psi^{-1}\left(\frac{\Lambda_k(t)}{F_\eta^{-1}(\exp(-\Lambda_{d+1}(t)))}\right)\right).\
\]
where $\Lambda_k(t),\Lambda_{d+1}(t)\sim \mathsf{G}(t,1)$. Here we require gamma processes for $d$ independent $\mathsf{Exp}(1)$ random variables and another for the common random variable $G$.

\begin{table}[h!]
\centering
\caption{Estimation of $\ell$ for a Clayton copula model}
\begin{tabular}{|c c c c c c|}

 \hline
 $\alpha$ & $\gamma$ & $\hat{\ell}^{IS}$ & $\hat{\ell}^{DS}$ & $RE^{IS}(\%)$ & $RE^{DS}(\%)$\\ [0.5ex] 
 \hline
 $0.95$ & $55$ & $0.0493$ & $0.0513$ & $1.83$ & $4.87$\\ 
 $0.99$ & $119$ & $0.0094$ & $0.0099$ & $4.14$ & $6.67$\\
 $0.995$ & $146$ & $0.0049$ & $0.0047$ & $5.49$ & $8.14$ \\ [1ex] 
 \hline
\end{tabular}

\label{table:6}
\end{table}

Once again, we can see that $\alpha+\hat{\ell}^{IS}$ and $\alpha+\hat{\ell}^{DS}$ equal to $1$ approximately. However, the relative error of the DS estimator is still larger than the relative error for the one-step IS estimator $RE^{IS}$. Thus, in this study the DS estimator does not perform as efficiently as the one-step IS estimator.

\section{Critical Analysis of the Dynamic Splitting Method}
In this section, we give a critical analysis of the proposed DS method when applied to rare-event probability estimation. The proposed DS method has advantages in terms of simplicity, versatility and simulation effort. Conversely, the DS approach can be inapplicable if the problem does not possess a quasi-monotonic importance function $S$. \newline
\textbf{Advantages:}
\begin{enumerate}
  \item Interpretability $-$ In the DS method, we estimate $\ell$ by computing the product of sequential conditional probabilities. Each conditional probability can be interpreted as a survival probability from the current time level to the next time level. This gives us a simple interpretation to our estimate of $\ell$ as a survival probability from the starting time to the terminal time.
  \item Versatility $-$ One of the biggest strengths of the proposed DS method is that it is versatile as it can generate any set of continuous random variables $\bX$ by using the inverse-transform method. In particular, we can generate sequential values of each variable such that $\{S(\bX)\geq \gamma\}$. This is possible as the monotonicity of the Gamma process is preserved from under the embedding transformation, thus allowing us generate the set $\{\bX:S(\bX)\geq \gamma\}$. We also note that unlike IS algorithms, DS does not require the computation of likelihood ratios and can be applied on any continuous pdf including heavy-tailed distributions. This is a major advantage for DS when compared to the one-step and two-step IS algorithms which are exclusively applicable to light-tailed distributions such as the Normal distribution. The same argument cannot hold against CE as the only main restriction in CE is that it only considers densities within the family of distributions which is not a strong assumption in itself.
  \item Simulation effort $-$ A natural question that would arise is the simulation effort and computational cost of DS since it simulates split paths of the Markov chain over many time levels. The danger of explosions can be avoided by a fixed effort implementation and paths $\bX$ for which $\{\bX:S(\bX)< \gamma\}$ are not simulated further.  We also note that unlike generalized splitting or particle methods, our proposed DS method does not require approximate MCMC sampling at each level of splitting. These observations ensure that the implementation of our proposed DS method would not be computationally expensive.
\end{enumerate}

\textbf{Limitations:}
\begin{enumerate}
\item Quasi-monotonicity $-$ Our proposed DS algorithm hinges on the existence of a quasi-monotonic importance function $S$. If the problem of interest has an importance function $S$ that does not possess this property then we cannot apply DS as the absence of quasi-monotonicity in $S$ implies we can no longer apply the embedding transformations to the problem and the decomposition of \ref{ell} into nested subsets no longer holds. Consequently, there is no connection between the static distributions of interest and a time-dependent Markov process, making DS inapplicable. 
\item Efficiency $-$ As shown by the results of numerical experiments earlier in this chapter, we can see that the RE of the DS method is much larger than current hallmark rare-event probability estimation algorithms such two-step IS and CE based algorithms \cite{chan2010efficient}. This would be attributed to the decomposition of time levels to estimate each conditional probability to be inadequate form of variance reduction in comparison to two-step IS and CE, both of which seek to obtain a density with reduced variance for efficient estimation of $\ell$.
\end{enumerate}

\chapter{Conclusions and Suggestions for Future Research}
\label{chapter: conclusion}
We have developed and illustrated a splitting method which is simple and effective to estimate rare event probabilities in the framework of popular copula credit risk models. This method is designed to estimate tail probabilities by decomposing the state space of the risk variables into nested subsets so that the rare event can be expressed as an intersection of these subsets. This decomposition allows us to achieve greater accuracy in estimating each conditional probability than estimating the rare probability itself. We have also shown that despite the inapplicability of the classical splitting method on static problems, one can always find an appropriate embedding transformation such that the static density being modelled can be taken as a snapshot of a continuous time Markov process at a particular instance in time. The illustrations of importance sampling estimators in this thesis, namely \emph{exponential twisting}, have relied on conditional independence of the model to simplify the problem to a sum of scaled Bernoulli random variables and attractive forms of the likelihood ratio. Our proposed modification to the dynamic splitting algorithm does not rely on conditional independence, rather it relies on the quasi-monotonicity property of the importance function. A large number of models either possess this property or can achieve it through a transformation. Hence, although the algorithm does not perform as efficiently as current hallmark importance sampling estimators, it is a versatile method that is applicable to any static problem for which random variables required in the model have prespecified distributions.

A possible direction for future research include developing more efficient adaptive dynamic splitting algorithms and variance reduction techniques to dynamic splitting, perhaps by applying exponential twisting to each candidate sample path that will be split. The latter is based on the observation that each sample path that is split in fact Binomial distributed with $s$ trials, that is $\mathsf{Bin}(s,c)$. Possible practical applications for future work may include financial and insurance risk management. This is motivated by the observation that the splitting framework can be interpreted as a model for survival  probabilities which is analogous to survival probabilities of insurance policyholders and insurers; and also default and bankruptcy rates of financial institutions. Specifically, it may be useful in the pricing of insurance products as well as risk and solvency capital allocation where the VaR and CVaR are possible risk measures.


\begin{appendices}
\chapter{Monte Carlo methods}
\section{One-step IS algorithm: Shift in Probabilities}
We simulate $L$ under the following model.
Let
\[
L=c_1 B_1+\cdots+c_m B_m= \v c^\top  \v B
\]
be the total loss incurred by a portfolio of $m$ obligors, where
$c_i$ is the loss incurred from the $i$-th obligor and $B_i\sim \mathsf{B}(P_i)$ is a Bernoulli random variable indicating whether the $i$-th obligor has defaulted. The distribution of the vector $\v B$ is implicitly
defined via 
\[
B_k\idef \bb I\{X_k>x_k\},\quad k=1,\ldots,m
\]
where $\{x_k\}$ are given fixed thresholds and $\v X$ has a continuous joint density $f(\v x)$. Thus, the likehihood  of the total loss $L$ is given by
\[
f_{L} (l) = \prod_{k=1}^m P_k^{b_k} (1-P_k)^{1-b_k}
\]

To simulate from the upper tail of the distribution, an exponential twist is applied to the probabilities $\{ P_i\}$ for $\theta \geq 0$ as the first step in importance sampling. It is defined as follows
\[
P_{k,\theta}=\frac{P_k e^{\theta c_k }}{1+P_k (e^{\theta c_k}-1)}.
\]

This change in probability measure results in a density that simulates higher values of $L$ with default indicators now generated by $B_i\sim \mathsf{B}(\tilde P_i)$. Note that the notation $L$ and $B_k$ have been preserved as the desired likelihood ratio should be described under the notation of the original distribution. As the probability of a particular default outcome is being compared across two densities, the likelihood ratio aims to describe the ratio of probabilities for the same outcome. The likelihood of $L$ under the exponentially twisted density $g_L$ is given by
\[
g_{ L} ( l) = \prod_{k=1}^m \tilde P_k^{ b_k} (1-\tilde P_k)^{1- b_k}.
\]

Hence the likelihood ratio $W$ is given by
\[
\begin{split}
W( l )  &= \frac{f_{ L}(l)}{g_{ L}(l)}\\
&= \prod_{k=1}^m \left(\frac{P_k}{\tilde P_k}\right)^{b_k} \left( \frac{1-P_k}{1-\tilde P_k} \right)^{1-b_k} \\
&=\prod_{k=1}^m \left(\frac{1+P_k (e^{\theta c_k}-1)}{e^{\theta c_k}}\right)^{b_k} (1+P_k (e^{\theta c_k}-1))^{1-b_k}\\
&=\prod_{k=1}^m \left(1+P_k (e^{\theta c_k}-1)\right) e^{-\theta b_k  c_k}\\
&=\exp\left(-\theta l + \sum_{k=1}^m \log\left(1+P_k(\exp(\theta c_k)-1)\right)\right)
\end{split}
\]

\section{Two-step IS algorithm: Shift in Factors}
As the second step in importance given in \cite{glasserman2005importance}, the mean of the factors $\v Z$ are shifted from $\v 0$ to $\v \mu$. Assuming at Normal copula, this means we now have $\v Z \sim\mathsf{N}(\v \mu,\m I_d)$ instead of $\v Z \sim\mathsf{N}(\v 0,\m I_d)$. Hence, the change in mean must also form part of the estimator. We note that the normalization constants of Normal densities are unchanged by a shift in the mean so the likelihood ratio is given by
\[
\frac{\exp{(-\frac{1}{2}\v Z^\top \v Z)}}{\exp{(-\frac{1}{2}\v {(\mu - Z)}^\top \v {(\mu - Z)})}}= \exp{(\frac{1}{2}\v \mu^\top \v \mu - \v \mu^\top \v Z)}
\]

\section{Archimedean copula sampling algorithm}
To verify that the algorithm in \cite{kroese2010cross} draws $\v U=(U_1, \cdots , U_d)$ from an Archimedean copula, we must show that
\[
P(\v U\leq \v u) = \psi^{-1}\left(\sum_{i=1}^d\psi(u_i)\right),
\]
where 
\[
(U_1,\ldots,U_d)=\left(\psi^{-1}\left(\frac{E_1}{\Lambda}\right),\ldots, \psi^{-1}\left(\frac{E_d}{\Lambda}\right)\right).
\]
The probability of $\v U$ can be written as
\[
\begin{split}
\bb P(\v U\leq \v u)&=\bb P( U_1\leq  u_1,\ldots,U_d\leq  u_d)\\
&=\bb P( E_1\geq  \Lambda\psi(u_1),\ldots,E_d\geq  \Lambda\psi(u_d))\quad \textrm{ since $\psi$ is invertible and decreasing}\\
&=\bb E_\Lambda\bb P( E_1\geq  \Lambda\psi(u_1),\ldots,E_d\geq  \Lambda\psi(u_d)|\Lambda = \lambda)\\
&=\bb E_\Lambda\prod_i\bb P( E_i\geq  \lambda\psi(u_i))\\
&=\bb E_\Lambda \prod_i\exp(-\lambda \psi(u_i))\\
&=\bb E_\Lambda\exp\left(-\lambda \sum_{i=1}^d\psi(u_i)\right)\\
&=\int_0^\infty \exp\left(-\lambda \sum_{i=1}^d\psi(u_i)\right) f_\Lambda(\lambda) \m d \lambda\\
&=\psi^{-1}\left(\sum_{i=1}^d\psi(u_i)\right),\qquad \textrm{as required}.
\end{split}
\]
\end{appendices}

\begin{appendices}
\chapter{Dynamic Splitting}
\section{Embedding transformations}
We now verify that the embedding transformations \eqref{embedtransform-dec} and \eqref{embedtransform-inc} do indeed have the desired distribution at $t=1$. For \eqref{embedtransform-dec} we have
\[
\begin{split}
\bb P (X_k(1)\leq x) &= \bb P (\exp(-\Lambda_k(1))\leq F_k(x))\\
&=\bb P (\Lambda_k(1)\geq -\log (F_k(x)))\\
&=\exp (\log (F_k(x)))\\
&= F_k(x).
\end{split}
\]
For \eqref{embedtransform-inc} we have
\[
\begin{split}
\bb P (X_k(1)\leq x) &= \bb P (1-\exp(-\Lambda_k(1))\leq F_k(x))\\
&=\bb P (\exp(-\Lambda_k(1))\geq 1-F_k(x))\\
&=\bb P (\Lambda_k(1)\leq -\log(1-F_k(x)))\\
&=1-\exp(\log(1-F_k(x)))\\
&= F_k(x).
\end{split}
\]
Hence both transformations yield the desired distribution $F_k(x)$ at $t=1$ as required.

\section{Ideal Case Analysis of Fixed Factor Splitting Algorithm}
\label{optimsplit}
We now present an analysis of the performance of the Fixed Factor Splitting algorithm under an ideal assumption. The assumption is that the  time levels $t_{i} \in \{t_1, \cdots ,t_{L} \}$ are selected such that the conditional probabilities $c=\bb P(S(\bX(t_i)) \gvn S(\bX(t_{i-1}))$  are exactly, rather than approximately, equal to $s$ for all $i$.

Let $N_i=|\scX_i|$ be the random number of states in the set $\scX_i=\{X(t_i):S(X(t_i))\geq \gamma\}$. At time $t_0=0$ we have $N_1=1$ as the algorithm begins with a single path for the Markov process $\bLam(t_1)$. If we denote the number of states in $\scX_{i+1}$ that are generated from the $j$-th state from $\scX_{i}$ by $Q_{j,i}$ then we have the branching process recursion
\[
N_{i+1} = Q_{1,i}+Q_{2,i}+\cdots+Q_{N_t,i}
\]
where it is clear that $Q_{j,i}\sim \mathsf{Bin}(s,c)$, that is, a Binomial distributed random variable with probability $c$. Thus, we have $\bb E [Q_{j,i}]=sc=1$ and $\Var(Q_{j,i})=sc(1-c)=1-c$. By standard branching process arguments \cite{harris2002theory} we have $\bb E[N_i]=1$ and $\Var (N_i)=(i-1)(1-c)$ for $1<i<L$. Hence, for the unbiased estimator $W=\frac{|\scX_L|}{s^{L-1}}=\frac{N_L}{s^{L-1}}$ we have $\bb E[W]=\ell=\frac{1}{s^{L-1}}$ and $\Var(W)=\frac{(L-1)(1-c)}{s^{2L-2}}$ with $\log(\ell)=(L-1)\log(s)$.

An estimator $\hat{\ell}$ of $\ell$ is \emph{logarithmically efficient} \cite{kroese2011handbook} if the following condition holds:
\[
\limsup_{\ell\downarrow 0} \left|\frac{\log(\Var(\hat{\ell}))}{\log(\ell^2)}\right|\geq 1.
\]
For the logarithmic efficiency criterion we have
\[
\begin{split}
\lim_{\ell\downarrow 0} \left|\frac{\log(\Var(W))}{\log(\ell^2)}\right|&=\lim_{\ell\downarrow 0} \left|\frac{\log(L-1)+\log(1-c)-(2L-2)\log(s)}{(2L-2)\log(s)}\right|\\
&=\lim_{s\uparrow \infty} \left|\frac{\log(L-1)+\log(1-c)-(2L-2)\log(s)}{(2L-2)\log(s)}\right|\\
&=1.
\end{split}
\]

Therefore, under the idealized assumption the estimator $W$ is logarithmically efficient. Note that the simulation effort, starting from $t_1$, is a random variable $s\sum_{i=1}^L N_i$ with expected value $s(L-1)$. The expected \emph{relative time variance product} \cite{kroese2011handbook} is thus given by
\[
\begin{split}
\frac{\Var(W)}{\ell^2}s(L-1)&=(L-1)(1-c)s(L-1)\\
&=(L-1)^2(s-1)\\
&=\left(\frac{\log\ell}{\log (s)}\right)^2(s-1),
\end{split}
\]
which is minimized as a function of $s$ for $s>1$ at $s=4.92155363$ or $s=5$ when constrained on the integers.

\section{Adaptive Dynamic Splitting}
The description of the dynamic splitting method described up to this point has allowed arbitrary choices for the intermediate time levels  $t_{i} \in \{t_1, \cdots ,t_{L-1} \}$. We now describe a pilot algorithm to select optimal values for the intermediate time levels $\{t_{i}\}$. The motivation behind this selection is to ensure that the conditional probabilities $\Pm(S(\bLam(t_{i}))\geq \gamma \gvn S(\bLam(t_{i-1}))\geq \gamma)$ are not too small and not rare-event probabilities so that they can be easily estimated with CMC. \cite{botev2012efficient} notes that this formulation will in fact lead to biased estimates of $\ell$ and complications in the computation of the variance of $\hat{\ell}$. As a consequence, the relative error of the estimator can be difficult to compute, thereby limiting comparisons and benchmarks against other Monte Carlo estimators. Thus, the \emph{Fixed Effort Splitting} algorithm is recommended as it will lead to unbiased estimates for $\ell$ and readily available estimates of variance by running the algorithm several times independently. The following algorithm implements a simple procedure similar to numerical root-finding by utilizing the Gamma Bridge Sampling algorithm.

\begin{algorithm}[H]
\caption{: Adaptive Dynamic Splitting}
\begin{algorithmic}
\REQUIRE {start time $t_s$; end time $t_e$; endpoints $\bLam(t_s)\leq \bLam(t_e)$; total sample at each level $s$; time tolerance $\epsilon_t$; proportion tolerance $\epsilon_p$}
\STATE{$\scT \leftarrow\emptyset$}
\STATE{$\bLam(t_l)\leftarrow \bLam(t_s)$}
\STATE{$\bLam(t_u)\leftarrow \bLam(t_e)$}
\STATE{$t_m\leftarrow \frac{t_s+t_e}{2}$}
\FOR{$j=1,\ldots,s$}
\STATE{Generate $\bLam_j(t_m)=(\Lambda_1(t_m),\ldots,\Lambda_{d}(t_m))^\top$ using Gamma Bridge Sampling with endpoints $\bLam(t_l)\leq \bLam(t_u)$}
\ENDFOR
\STATE{$i\leftarrow 1$}
\WHILE{$t_e-t_m>\epsilon_t$}
\WHILE{$|\frac{1}{s}\sum_{j=1}^s \bb I\{S(\bLam_j(t_m))\geq \gamma\}-\rho|>\epsilon_p$}
\IF{$\frac{1}{s}\sum_{j=1}^s \bb I\{S(\bLam(t_m)\geq \gamma\}-\rho>0$}
  \STATE{$\bLam(t_l)\leftarrow \bLam(t_m)$}
  \STATE{$t_m \leftarrow \frac{t_m+t_u}{2}$} 
  \ELSIF{$\frac{1}{s}\sum_{j=1}^s \bb I\{S(\bLam(t_m)\geq \gamma\}-\rho<0$}
  \STATE{$\bLam(t_u)\leftarrow \bLam(t_m)$}
  \STATE{$t_m \leftarrow \frac{t_l+t_m}{2}$} 
  \ENDIF
  \FOR{$j=1,\ldots,s$}
\STATE{Generate $\bLam_j(t_m)=(\Lambda_1(t_m),\ldots,\Lambda_{d}(t_m))^\top$ using Gamma Bridge Sampling with endpoints $\bLam(t_l)\leq \bLam(t_u)$}
\ENDFOR
\ENDWHILE
\STATE{$t_i\leftarrow t_m$}
\STATE{add $t_i$ to $\scT$}
\STATE{$i\leftarrow i+1$}
\ENDWHILE
 \RETURN{$\scT$}
\end{algorithmic}
\end{algorithm}

\begin{algorithm}[H]
\caption{: Gamma bridge sampling}
\begin{algorithmic}
\REQUIRE {Endpoints $\bLam(t_l)\leq \bLam(t_u)$ and  $t\in(t_l,t_u)$.}
\FOR{$k=1,\ldots,d$}
\STATE{$B_k\sim \mathsf{Beta}(t-t_l,t_u-t)$, independently}
\STATE{$\Lambda_k(t)\g\Lambda_k(t_l)+ (\Lambda_k(t_u)-\Lambda_k(t_l))B_k$}
\ENDFOR
 \RETURN{$\bLam(t)=(\Lambda_1(t),\ldots,\Lambda_d(t))^\top$.}
\end{algorithmic}
\end{algorithm}
\end{appendices}

\clearpage																			
\addcontentsline{toc}{chapter}{References}      
\bibliographystyle{plain}
\bibliography{main}

\begin{thebibliography}{10}

\bibitem{asmussen2007stochastic}
S{\o}ren Asmussen and Peter~W Glynn.
\newblock {\em Stochastic simulation: Algorithms and analysis}, volume~57.
\newblock Springer Science \& Business Media, 2007.

\bibitem{au2001estimation}
Siu-Kui Au and James~L. Beck.
\newblock Estimation of small failure probabilities in high dimensions by
  subset simulation.
\newblock {\em Probabilistic Engineering Mechanics}, 16(4):263--277, 2001.

\bibitem{barndorff2001multivariate}
Ole~E. Barndorff-Nielsen, Jan Pedersen, and Ken-iti Sato.
\newblock Multivariate subordination, self-decomposability and stability.
\newblock {\em Advances in Applied Probability}, pages 160--187, 2001.

\bibitem{bassamboo2008portfolio}
Achal Bassamboo, Sandeep Juneja, and Assaf Zeevi.
\newblock Portfolio credit risk with extremal dependence: Asymptotic analysis
  and efficient simulation.
\newblock {\em Operations Research}, 56(3):593--606, 2008.

\bibitem{bluhm2010introduction}
Christian Bluhm, Ludger Overbeck, and Christoph Wagner.
\newblock {\em Introduction to credit risk modeling}.
\newblock Crc Press, 2010.

\bibitem{botev2008efficient}
Zdravko~I. Botev and Dirk~P. Kroese.
\newblock An efficient algorithm for rare-event probability estimation,
  combinatorial optimization, and counting.
\newblock {\em Methodology and Computing in Applied Probability},
  10(4):471--505, 2008.

\bibitem{botev2012efficient}
Zdravko~I. Botev and Dirk~P. Kroese.
\newblock Efficient monte carlo simulation via the generalized splitting
  method.
\newblock {\em Statistics and Computing}, 22(1):1--16, 2012.

\bibitem{brereton2013monte}
Tim~J Brereton, Dirk~P Kroese, and Joshua~C Chan.
\newblock Monte carlo methods for portfolio credit risk.
\newblock 2013.

\bibitem{brooks2011handbook}
Steve Brooks, Andrew Gelman, Galin Jones, and Xiao-Li Meng.
\newblock {\em Handbook of Markov Chain Monte Carlo}.
\newblock CRC press, 2011.

\bibitem{bucklew2013introduction}
James Bucklew.
\newblock {\em Introduction to rare event simulation}.
\newblock Springer Science \& Business Media, 2013.

\bibitem{chan2010efficient}
Joshua~CC Chan and Dirk~P Kroese.
\newblock Efficient estimation of large portfolio loss probabilities in
  t-copula models.
\newblock {\em European Journal of Operational Research}, 205(2):361--367,
  2010.

\bibitem{dembo2009large}
Amir Dembo and Ofer Zeitouni.
\newblock {\em Large deviations techniques and applications}, volume~38.
\newblock Springer Science \& Business Media, 2009.

\bibitem{frey2003dependent}
R{\"u}diger Frey and Alexander~J McNeil.
\newblock Dependent defaults in models of portfolio credit risk.
\newblock {\em Journal of Risk}, 6:59--92, 2003.

\bibitem{glasserman2003monte}
Paul Glasserman.
\newblock {\em Monte Carlo methods in financial engineering}, volume~53.
\newblock Springer Science \& Business Media, 2003.

\bibitem{glasserman1999asymptotically}
Paul Glasserman, Philip Heidelberger, and Perwez Shahabuddin.
\newblock Asymptotically optimal importance sampling and stratification for
  pricing path-dependent options.
\newblock {\em Mathematical finance}, 9(2):117--152, 1999.

\bibitem{glasserman1998large}
Paul Glasserman, Philip Heidelberger, Perwez Shahabuddin, and Tim Zajic.
\newblock A large deviations perspective on the efficiency of multilevel
  splitting.
\newblock {\em Automatic Control, IEEE Transactions on}, 43(12):1666--1679,
  1998.

\bibitem{glasserman1998look}
Paul Glasserman, Philip Heidelberger, Perwez Shahabuddin, and Tim Zajic.
\newblock {\em A look at multilevel splitting}.
\newblock Springer, 1998.

\bibitem{glasserman2005importance}
Paul Glasserman and Jingyi Li.
\newblock Importance sampling for portfolio credit risk.
\newblock {\em Management science}, 51(11):1643--1656, 2005.

\bibitem{glynn1996importance}
Peter~W Glynn.
\newblock Importance sampling for monte carlo estimation of quantiles.
\newblock In {\em Mathematical Methods in Stochastic Simulation and
  Experimental Design: Proceedings of the 2nd St. Petersburg Workshop on
  Simulation}, pages 180--185, 1996.

\bibitem{harris2002theory}
Theodore~E Harris.
\newblock {\em The theory of branching processes}.
\newblock Courier Corporation, 2002.

\bibitem{hong2014review}
L.~Jeff Hong, Zhaolin Hu, and Guangwu Liu.
\newblock Monte carlo methods for value-at-risk and conditional value-at-risk:
  A review.
\newblock {\em ACM Transactions on Modeling and Comput. Simulation},
  24(4):22:1--22:37, 2014.

\bibitem{hong2011monte}
L.~Jeff Hong and Guangwu Liu.
\newblock Monte carlo estimation of value-at-risk, conditional value-at-risk
  and their sensitivities.
\newblock In {\em Proceedings of the Winter Simulation Conference}, pages
  95--107, 2011.

\bibitem{kahn1951estimation}
Herman Kahn and Ted~E. Harris.
\newblock Estimation of particle transmission by random sampling.
\newblock {\em National Bureau of Standards applied mathematics series},
  12:27--30, 1951.

\bibitem{kang2005fast}
Wanmo Kang and Perwez Shahabuddin.
\newblock Fast simulation for multifactor portfolio credit risk in the t-copula
  model.
\newblock In {\em Proceedings of the 37th conference on Winter simulation},
  pages 1859--1868, 2005.

\bibitem{kroese2010cross}
Dirk~P. Kroese, Reuven~Y. Rubinstein, and Peter~W. Glynn.
\newblock The cross-entropy method for estimation.
\newblock {\em Handbook of Statistics, edited by V. Govindaraju and CR Rao},
  31, 2010.

\bibitem{kroese2011handbook}
Dirk~P. Kroese, Thomas Taimre, and Zdravko~I. Botev.
\newblock {\em Handbook of Monte Carlo Methods}, volume 706.
\newblock John Wiley \& Sons, 2011.

\bibitem{li1999default}
David~X Li.
\newblock On default correlation: A copula function approach.
\newblock {\em Available at SSRN 187289}, 1999.

\bibitem{rubinstein2011simulation}
Reuven~Y Rubinstein and Dirk~P Kroese.
\newblock {\em Simulation and the Monte Carlo method}, volume 707.
\newblock John Wiley \& Sons, 2011.

\bibitem{rubinstein2013cross}
Reuven~Y Rubinstein and Dirk~P Kroese.
\newblock {\em The cross-entropy method: a unified approach to combinatorial
  optimization, Monte-Carlo simulation and machine learning}.
\newblock Springer Science \& Business Media, 2013.

\bibitem{scott2015extensions}
Alexandre Scott.
\newblock Extensions of the cross-entropy method with applications to diffusion
  processes and portfolio losses.
\newblock 2015.

\bibitem{scott2015general}
Alexandre Scott and Adam Metzler.
\newblock A general importance sampling algorithm for estimating portfolio loss
  probabilities in linear factor models.
\newblock {\em Available at SSRN}, 2015.

\bibitem{van2000asymptotic}
Aad~W Van~der Vaart.
\newblock {\em Asymptotic statistics}, volume~3.
\newblock Cambridge university press, 2000.

\end{thebibliography}

\end{document}